\documentclass[twocolumn]{aastex63}

\usepackage{amsmath}
\received{June 1, 2019}
\revised{January 10, 2019}
\accepted{\today}
\submitjournal{ApJ}

\shorttitle{Br$\gamma$-bar in the GC}
\shortauthors{Pei$\beta$ker et al.}
\begin{document}

\title{Near- and mid-infrared observations in the inner tenth of a parsec of the Galactic center \\
\small Detection of proper motion of a filament very close to Sgr~A*}

\correspondingauthor{Florian Pei{\ss}ker}
\email{peissker@ph1.uni-koeln.de}
%\uccode ß="1E9E
%\author[0000-0002-0786-7307]{Florian Pei{\ss}ker}
%\author[0000-0002-9850-2708]{Florian Pei{"1E9E}ker}
\author[0000-0002-9850-2708]{Florian Pei$\beta$ker}
\affil{I.Physikalisches Institut der Universit\"at zu K\"oln, Z\"ulpicher Str. 77, 50937 K\"oln, Germany}
\author[0000-0001-6049-3132]{Andreas Eckart}
\affil{I.Physikalisches Institut der Universit\"at zu K\"oln, Z\"ulpicher Str. 77, 50937 K\"oln, Germany}
\affil{Max-Plank-Institut f\"ur Radioastronomie, Auf dem H\"ugel 69, 53121 Bonn, Germany}
\author[0000-0001-7134-9005]{Nadeen B. Sabha}
\affil{Institut f\"ur Astro- und Teilchenphysik, Universit\"at Innsbruck, Technikerstr. 25, A-6020 Innsbruck, Austria}
\affil{I.Physikalisches Institut der Universit\"at zu K\"oln, Z\"ulpicher Str. 77, 50937 K\"oln, Germany}
\author[0000-0001-6450-1187]{Michal Zaja\v{c}ek}
\affil{Center for Theoretical Physics, Polish Academy of Sciences, Al. Lotników 32/46, 02-668 Warsaw, Poland}
\affil{I.Physikalisches Institut der Universit\"at zu K\"oln, Z\"ulpicher Str. 77, 50937 K\"oln, Germany}
\author[0000-0002-4408-0650]{Harshitha Bhat}
\affil{I.Physikalisches Institut der Universit\"at zu K\"oln, Z\"ulpicher Str. 77, 50937 K\"oln, Germany}



\begin{abstract}
We analyze the gas and dust emission in the immediate vicinity of the supermassive black hole Sgr~A* at the Galactic center (GC) with the ESO VLT (Paranal/Chile) instruments SINFONI and VISIR. The SINFONI H+K data cubes show several emission lines with related line map counterparts. From these lines, the Br$\gamma$ emission is the most prominent one and appears to be shaped as a bar extending along the North-South direction.
%The detected bar-like line emission feature in the field of view is stretched in North-South direction.
With VISIR, we find a dusty counterpart to this filamentary emission. In this work, we present evidence that this feature can be most likely connected to the mini-spiral and potentially influenced by the winds of the massive stars in the central cluster or an accretion wind from Sgr~A*. 
To this end, we co-add the SINFONI data between 2005 and 2015. The spectroscopic analysis reveals a range of Doppler-shifted emission lines. We also detect substructures in the shape of clumps that can be investigated in the channel maps of the Br$\gamma$-bar. In addition, we compare the detection of the near-infrared (NIR) Br$\gamma$ feature to PAH1 mid-infrared (MIR) observations and published 226 GHz radio data.
These clumps show a proper motion of about $320$km/s that are consistent with other infrared continuum detected filaments in the Galactic center. Deriving a mass of
$2.5\,\times\,10^{-5}\,M_{\odot}$  for the investigated Br$\gamma$-feature shows an agreement with former derived masses for similar objects. Besides the North-South Br$\gamma$-bar, we find a comparable additional East-West feature. Also, we identify several gas reservoirs that are located west of Sgr~A* that may harbor dusty objects.
\end{abstract}



\section{Introduction}
Investigating the vicinity of supermassive black holes (SMBHs), and in particular the vicinity of the SMBH associated with the compact radio source Sgr~A*, is often connected to the investigation of gas and dust on large ($>>$ 1 pc) and small scales ($<$ 1 pc ). This is important if one wants to characterize star formation or to determine the mass of the SMBH itself. In this context, \cite{Wollman1977} used the mini-spiral gas velocities to derive the Sgr~A* SMBH mass. \cite{Yusef-Zadeh1998} and \cite{Zhao1998} also observed proper motions of gaseous-dusty structures and filaments based on their radio continuum detections in the vicinity of Sgr~A* and the Galactic Center (GC) stellar association IRS13E. \cite{muzic2007} completed that picture by expanding the analysis towards shorter wavelengths in the near-infrared (NIR). These publications derive typical gas and dust filament velocities of several hundred km/s. 

However, extragalactic studies of gas and dust on larger scales by \cite{boeker2008} 
 lead to the introduction of two models that could describe the feeding process of SMBHs as well as the provision of a gas reservoir which is responsible for star formation -- ``popcorn model'' and ``pearls on a string'' scenarios. A large amount of the dusty and gaseous material typically accumulates along the ring on the scale of 100 parsecs and less at the innermost dynamical resonance referred to as the inner or nuclear Lindblad resonance \citep{1998MNRAS.295..463F}. \citet{boeker2008} discuss two scenarios of star-formation along the nuclear ring. The ``popcorn model'' describes a scenario when the gas is accumulated along the ring rather uniformly so that the critical density is reached throughout the whole ring. As a result, there is no age gradient of star-forming clusters along the ring. The ``pearls on a string'' model operates with the localized burst of star-formation in so-called overdensity regions, which can often be associated with the sites where the gas material enters the ring. The star-forming clusters are then formed along the ring in a sequential way which leads to the age gradient. The ``pearls on a string'' scenario is supported by observational data \citep{boeker2008,2019A&A...622A.128F}.

 For the Galactic center, \cite{Nayakshin2007} and \cite{Jalali2014} modeled the gravitational influence of the SMBH on molecular clouds on intermediate scales that are moving from a distance of several parsecs towards Sgr~A*. The authors find indications that in-falling material could trigger star formation in the close vicinity of the SMBH. On larger scales, this could be linked to the proposed ''Pearls on a string" model. It is evident, that the investigation of gas and dust reservoirs located in the inner and outer parsecs contributes to the understanding of star formation \citep{boeker2008, Jalali2014} and black hole feeding.
 
  A key question is how the material is transported from the larger scales all the way to Sgr~A*. An occasional collision of clumps inside the circumnuclear disk (CND), which is located between $1.5$ to $7$ pc, can lead to the loss of angular momentum and the formation of streamers that are tidally interacting with the SMBH -- the Minispiral streamers can be the manifestation of this process \citep{Jalali2014}. Another way to fill the central cavity inside the CND is the interaction of stellar winds with the inner rim of the CND, which can also lead to the formation of infalling denser clumps that are tidally interacting with Sgr~A* \citep{2016MNRAS.459.1721B}. Finally, thermal instability could have operated in the central parsec during the phases of enhanced activity of Sgr~A*, which would create a multiphase medium where the warm and more diluted plasma material is in approximate pressure equilibrium with the colder and denser gas \citep{2014MNRAS.445.4385R,2017MNRAS.464.2090R}.

In this work, we investigate the Br$\gamma$ gas reservoir in the direct vicinity of Sgr~A*, the SMBH of our host Galaxy. We use H+K data obtained with SINFONI, a Very Large Telescope (VLT) instrument which is now decommissioned\footnote{SINFONI was decommissioned end of June 2019.}. We investigate the ionized gas located close to the line of sight towards the S-cluster ($\sim\,$0.04 pc) in the channel maps of the related data-cubes. Additionally, we connect our detections to the 226 GHz radio observations executed by \cite{Yusef-Zadeh2017-ALMAVLA} and mid-infrared (MIR) VISIR data from 2018 to investigate the gas filaments at longer wavelength regimes.\newline
In section \ref{sec:data} we shortly introduce the methods used for the observations and the analysis of the obtained data. This is followed by section \ref{sec:results} where we present the results. These results are then discussed in section \ref{sec:discussion}. In section \ref{sec:conclusions}, we summarize our findings and present our conclusions.

\section{Data $\&$ Observations}
\label{sec:data}

Here, we briefly describe the observations, the data reduction, and the methods that are used for the data analysis.

\subsection{SINFONI}

The VLT Spectrograph for INtegral Field Observations in the Near Infrared (SINFONI) data presented in this work had already been used for \cite{Valencia-S.2015}, \cite{Peissker2019}, \cite{peissker2020a}, and \cite{peissker2020b} and are described there in detail. To provide a complete picture about the analysis, the data is listed in Appendix \ref{sec:data_used}. The observations were carried out at the UT4/VLT (Paranal, Chile). For the spectroscopic analysis, we use data-cubes between 2005 and 2015 centered on Sgr~A* to ensure an adequate signal to noise ratio. For determining the proper motion, we include also the data cube of 2018 to increase the baseline for the analysis.
The position of Sgr~A* is determined through its offset from the star S2 for which we know the
orbital elements very well \citep[see][]{Parsa2017, Gravity2019}. Here, in particular, we investigate the line emission of the Br$\gamma$ gas distribution and the channel maps of the SINFONI H+K data-cubes.
An image of the Br$\gamma$ line flux and the related H+K-band spectrum are shown in Fig.\ref{fig:bar}.

\subsection{VISIR}

The mid-infrared spectrometer and imager (VISIR) is mounted at the ESO UT2 (UT3 at the time of the observation). We used the smallest field of view (FOV) with a spatial pixel size of 0".045 and a total image size of 38".0 x 38".0 per image in the N-band PAH1 filter ($8.59\,\mu m$). To increase the signal to noise (S/N), we shift and add the single images to create a 40".0 x 40".0 overview image. We used the standard calibration files that are provided within the ESO pipeline. To suppress the background, differential observations with the chop/nod mode of VISIR are executed. This allows tracing faint point sources and in particular extended features at this wavelength. The observation was executed in 2016\footnote{Program ID: 097.C-0023} and is part of a larger survey (Sabha et al, in prep). From this survey, we choose the data with the best detectability of the Br$\gamma$-features.
%Dark and Flat-field frames are not provided after the instrumental update in 2012.

\subsection{Low-pass filter}

We discuss the need for different filter techniques in \cite{Peissker2019, peissker2020a, peissker2020b}. For the VISIR data, applying a low pass filter is an appropriate choice as it eliminates the influence of overlapping PSFs and preserves the shape of extended and elongated structures. With this technique we get access to information that would be suppressed due to the crowded field of view in the central few arcseconds of the GC. In order to obtain high angular resolution information %(as shown in the zoomed portion of Fig.\ref{})
we smooth the input file with a Gaussian (normalized to unity flux) that matches the size of the PSF of the data. This smoothed image is then subtracted from the input file. After removing negativities, the resulting image is smoothed again with a Gaussian. Depending on the required angular resolution and sensitivity as defined by the scientific goal, the size of the smoothing Gaussian can be adjusted.

\section{Results}
\label{sec:results}
In this section, we will show the results of the GC observations with SINFONI and VISIR. The bar-like structure of the Br$\gamma$ distribution and additional gas reservoirs can be observed in the SINFONI H+K channel maps and with the VISIR PAH1 filter. We compare these features to structures detected in the 226 GHz radio domain by \cite{Yusef-Zadeh2017}. %We determine the position, the spectral properties, and the proper motion of the Br$\gamma$-bar.
%In Fig.\ref{fig:bar2005} we compare the Br$\gamma$ line flux distribution to structures detected in the 226 GHz radio domain by \cite{Yusef-Zadeh2017}.
%In Fig.\ref{fig:mir_overview} we compare the Br$\gamma$ line emission with  the continuum emission in the MIR (see below).

\subsection{Spectral emission lines}
\label{sec:spectral_emission_lines}
\begin{figure*}[ht!]
	\centering
	\includegraphics[width=1.\textwidth]{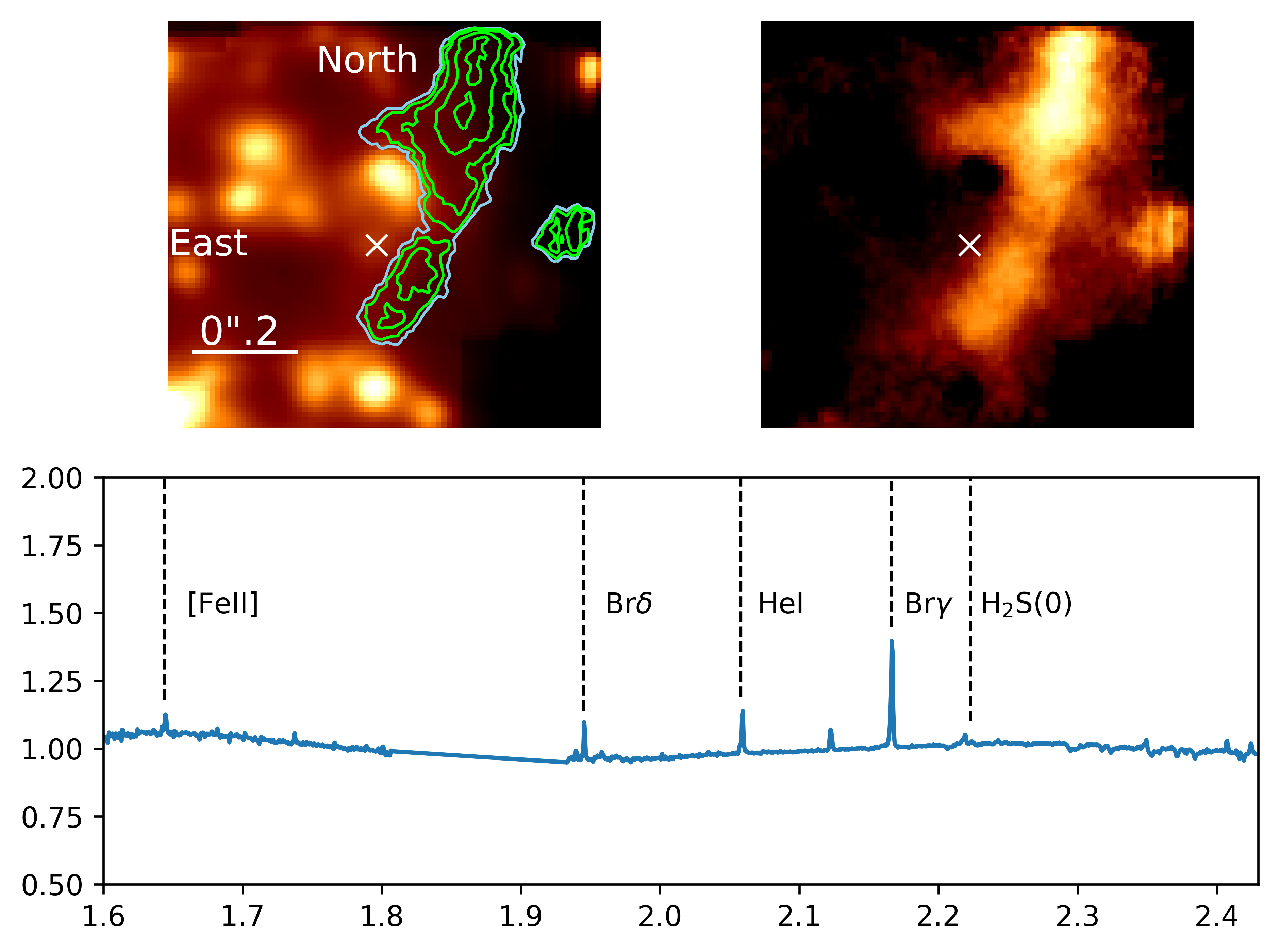}
	\caption{Combined SINFONI data-cubes between 2005 and 2015. The cubes are centered on the position of Sgr~A* (marked with a white x). The upper left image shows the K-band continuum counterpart of the related data-cube. The green contours refer to the Br$\gamma$ line map emission shown in the upper right image. We set the levels of the green contours to 52.5$\%$, 56.5$\%$, 65$\%$, and 75$\%$. The light-blue contour line in the upper left image is set to 50$\%$.} The lower panel shows the spectrum for the Br$\gamma$-bar presented in the upper right image. The dashed line marks the rest-wavelength of the related emission. Another feature similar to the Br$\gamma$-bar can be seen at the right edge of the FOV (see upper left image and Sec. \ref{sec:2005_L0}). The telluric emission between $1.80\,-\,1.93\mu m$ is flattened. 
The subtracted continuum emission of S2 in the upper right image results in a (an artificial)  footprint in the investigated Br$\gamma$-bar. We note, that the croissant shape of S2 just above Sgr~A* is an outcome of the co-adding.
\label{fig:bar}
\end{figure*}

Here, we present the results of the spectral line analysis of the bar which is shown in Fig. \ref{fig:bar}. We spatially select the feature in the data-cube for which its spectrum shows [FeII], Br$\gamma$, Br$\delta$, He$I$, H$_2$S(1), H$_2$S(0), some CO features, and H$_2$Q(1) line emission.
Since not every line reveals a strong counterpart image in its emission in the corresponding channel maps, we emphasize the investigation of the Br$\gamma$, Br$\delta$, HeI, [FeII], and H$_2$S(0) emission (see Table \ref{tab:emissionlines}). These lines show clearly identifiable flux distributions and can be safely related to  
ionized emission of a gaseous bar (see Fig. \ref{fig:line_overview}). An outliers seems to be the blue-shifted H$_2$S(0) emission line. Even though we detect an extended related H$_2$S(0) line emission that can be clearly connected to the Br$\gamma$-bar, the Doppler-shift of this distribution indicates the presence and contribution of embedded and/or background/foreground objects (see the discussion in Sec. \ref{sec:H2_line}). For the sake of completeness, the line is still listed in Table \ref{tab:emissionlines}.\newline
\begin{table*}[hbt!]
    \centering
    \begin{tabular}{lcr}
         \hline %& flux [10$^{-16}$ erg/s/cm$^{2}$]
         \hline
           Spectral line ($@$rest wavelength [$\mu$m]) & central wavelength [$\mu$m] & Velocity [km/s] \\
         \hline
         
          [FeII] $@$1.64400 $\mu$m     & 1.64476 $\pm$ 0.00022 &  138.68 $\pm$ 40.14 \\
          Br$\delta$ $@$1.94509 $\mu$m & 1.94548 $\pm$ 0.00010 &   60.15 $\pm$ 15.42 \\
          HeI $@$2.05869 $\mu$m        & 2.05951 $\pm$ 0.00017 &  119.49 $\pm$ 24.77 \\
          Br$\gamma$ $@$2.16612 $\mu$m & 2.16650 $\pm$ 0.00010 &   52.62 $\pm$ 15.42 \\
          H$_2$S(0) $@$2.22329 $\mu$m  & 2.21923 $\pm$ 0.00030 & -547.83 $\pm$ 40.48 \\
       \hline
    \end{tabular}
    \caption{Emission lines of the Br$\gamma$-bar based on the combined SINFONI data-cube combining data from observing periods 
obtained within a period of almost 10 years. The related spectrum is shown in Fig. \ref{fig:bar}.}
    \label{tab:emissionlines}
\end{table*}
We find, that the hydrogen recombination lines lie at a velocity of about 50-60 km/s and the lines that stand for a higher excitation ([FeII] and HeI) are at a higher velocity of a about 120-140 km/s. This may indicate that the hydrogen lines trace the bulk of our bar feature where as the higher excitation
lines may be linked to a wind phenomenon (acting at its surface) resulting in a slightly higher redshift.

For the HeI/Br$\gamma$ ratio we find a value of 0.7 which is in agreement with the ratios of (0.35 - 0.7) observed in the GC \citep{Gillessen2012}. 
Compared to the surrounding Br$\gamma$ emission extended on larger scales, the intensity of the line at the position of the bar and especially of the clump-like enhancements therein varies by around $10\%$. We derive therefore a HeI/Br$\gamma$ ratio of $0.7\,\pm\,0.1$.

The extended Br$\gamma$ line emission as well as the MIR PAH1 continuum image (see Sec. \ref{sec:MIR}) of the central few arcseconds clearly show a North-South as well as an East-West feature. Both contain substructures which are discussed in the following sections.

\subsection{Substructures of the North-South Br$\gamma$-bar}
In Fig. \ref{fig:bar}, we show a K-band image and a Br$\gamma$ channel map that is extracted from the combined data-cube that covers data sets obtained over 
a period of almost 10 years. 
The expected smearing effect due to proper motions of the features 
is significantly smaller than their extent and is eliminated by centering the individual data-cubes on the position of Sgr~A*.
Around 400 data-cubes with a single exposure time between 400-600s are used.
With the clear detection of clumps (in the following indexed by $cl$) in the extended Br$\gamma$-features, we can determine the filling factor $f$ of these components as derived by
\begin{equation}
\label{eq:filling}
    f\,=\,f_{\infty}\,+\,(1.0-f_{\infty})\exp\Big(-\frac{v}{v_{\rm cl}} \Big) \,\,\,\,.
\end{equation}
This equation is introduced by \cite{Hillier2003}. The authors use it to derive the filling factor for clumps that are created in the winds of two O-stars with a shared age of around 4.4 Myr in the Small Magellanic Cloud (SMC). 
Strong winds also prevail in the central region of the GC stellar cluster
which justifies in combination with the recent publication by \cite{Calderon2020} the use of Eq. \ref{eq:filling}. An effective use of the filling factor is described by \cite{Yusef-Zadeh2013} where the authors derive the number of clumps via $n_{cl}\,=\,f/(4/3 \pi r^{3})$. For $f=1$ there would be no clumping.
It is obvious, that a filling factor $f_{\infty}$ for the Br$\gamma$-bar far away from the observed emission is not equal to 1, but rather at lower values.
 Therefore, we assume an upper limit for $f_{\infty\,up}=$0.9 and a lower limit of $f_{\infty\,low}=$0.1 for equation Eq. \ref{eq:filling}. For $n_{cl}\,=\,2-4$ we derive values for $f$ close to 0.5.

The derived velocity $v_{cl}$ of the clumps in the Br$\gamma$-bar are based on the line-of-sight (LOS) velocity presented in Table \ref{tab:emissionlines}. \cite{muzic2010} state that the wind velocity $v$, prevailing in the S-cluster region, is about 750 km/s which is consistent with wind values derived by \cite{Najarro1994}. With that, we derive an average filling factor of $f_{avg}\, = \, 0.5$. From \cite{Gillessen2012}, we use for the case B recombination of Br$\gamma$
\begin{align}
    \rho_{{\rm cl}} & =2.6\times10^{5}\left(\frac{f}{0.5}\right)^{-\frac{1}{2}}\left(\frac{R}{3\,{\rm mpc}}\right)^{-\frac{3}{2}}\times \notag \\
    & \times \left(\frac{T}{10^4\,{\rm K}}\right)^{0.54}{\rm cm^{-3}}
    \label{eq_bar_clump_density}
\end{align}
where $\rho_{cl}$ donates the clump density. Following the analysis of \cite{Gillessen2012}, we adapt the assumed electron temperature of $10^{4}$K \citep[see also][]{Zhao2009}. With a radius of 75 mas/clump ($\hat{=}$ 3 mpc/clump) in the Br$\gamma$-bar, we get a mass per clump of
\begin{align}
    M_{\rm cl}& =1.7\times10^{28}\left(\frac{f}{0.5}\right)^{-\frac{1}{2}}\left(\frac{R}{3\,{\rm mpc}}\right)^{-\frac{3}{2}}\times \notag \\
    &\times \left(\frac{T}{10^4\,{\rm  K}}\right)^{0.54}{\rm g\,/cl}\,
     \approx\,2.1\,M_{\oplus}\,/{\rm cl} \,\,\,\, .
\end{align}
Since we detect around 4 individual clumps with comparable properties in the SINFONI FOV, we derive a lower limit for the mass of the observed Br$\gamma$-bar clumps of $8.4\,M_{\oplus}$ or $2.5\,\times\,10^{-5}\,M_{\odot}$. From the HeI/Br$\gamma$ ratio $\eta$, we get $\eta\,\approx\,0.7$. The Br$\gamma$ emission feature is obviously not homogeneous as can be seen in Fig. \ref{fig:bar}. In order for the presumably thin material to reproduce the line ratio $\eta$, the former derived volume filling factor of $\leq\,0.5$ is justified.

%{\bf remove:
%As mentioned in the caption of Fig. \ref{fig:bar}, similar features to the Br$\gamma$-bar can be observed. We show this %features in Fig. \ref{fig:bar2005}. For example, L1 shows an angle difference to L0 of $\Delta\,=\,60^{\circ}$. However, %the SINFONI data, that covers the area shown in Fig. \ref{fig:bar2005} is limited. Therefore, additional data that covers %the region west of Sgr~A*, is needed (e.g. MIRI/NEARSPEC data from the JWST).
%}

\subsection{Substructures of the East-West Br$\gamma$-bar and additional gas reservoirs}
\label{sec:2005_L0}
\begin{figure*}[ht!]
	\centering
	\includegraphics[width=.75\textwidth]{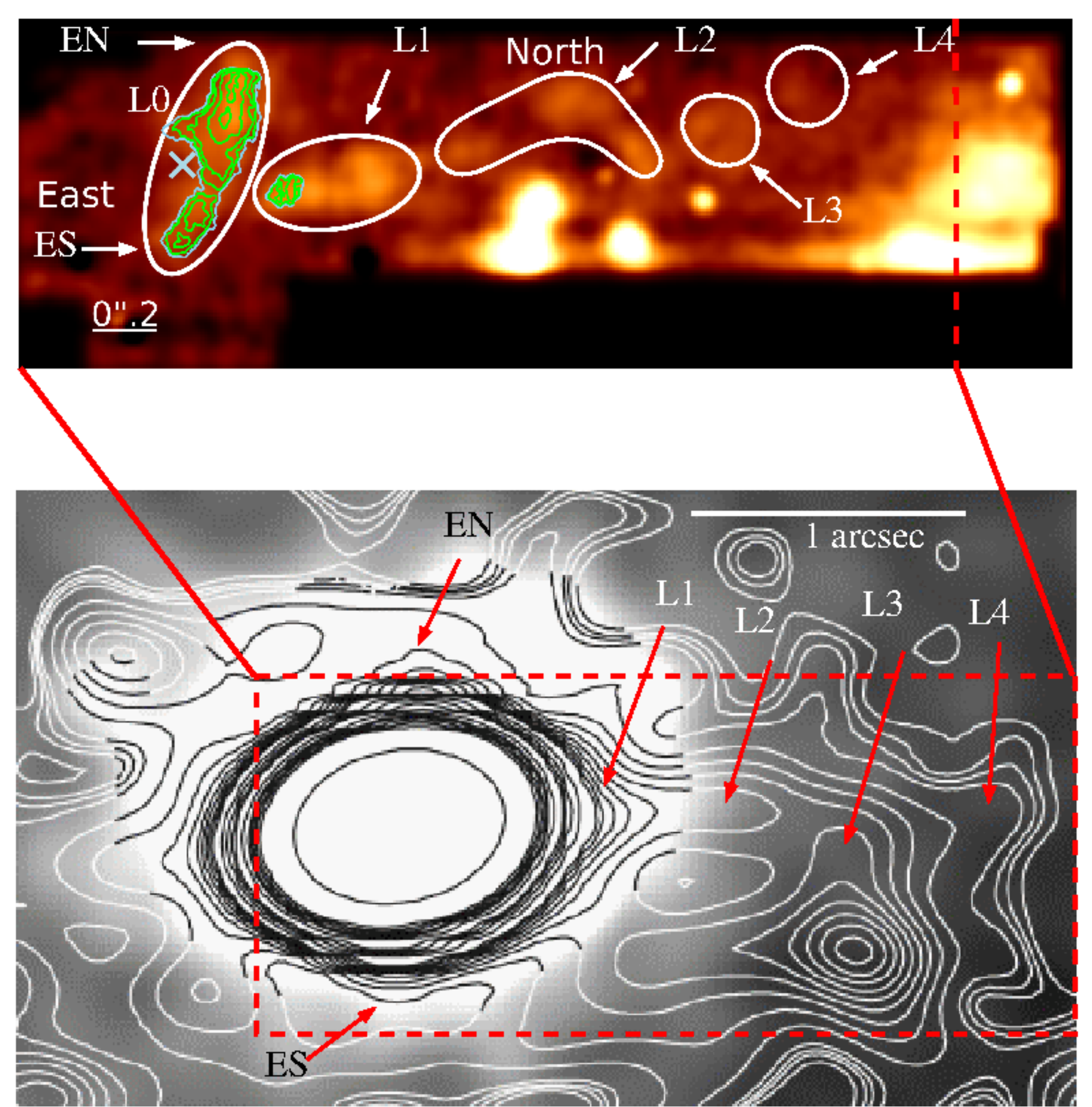}
	\caption{Galactic center in 2005. The upper panel shows a Br$\gamma$ line map of the SINFONI data cube smoothed with a Gaussian of 6 px. West of the north-south Br$\gamma$-bar (L0, the upper edge of the bar is marked with EN, the lower edgde with ES) we observe several gas reservoirs which are marked with arrows and white color. The estimated position of Sgr~A* is indicated by a light-blue cross. The contour lines are adapted from Fig. \ref{fig:bar}.
In the lower panel we show the same FOV in the 226 GHz radio domain as presented by \cite{Yusef-Zadeh2017}. We identify individual radio structures with source components detected at the positions of the Br$\gamma$ line flux in the infrared (see the upper panel)}.
\label{fig:bar2005}
\end{figure*}

Only the SINFONI data set of 2005 covers the area west of Sgr~A*. Because of the limited data quality, we smooth the Br$\gamma$ line map with a 6 px Gaussian. 
Besides the Br$\gamma$-bar, we find several other gas reservoirs west of Sgr~A* in different shapes. We name the detected features L0-L4 (see Fig. \ref{fig:bar2005}).
Since these gas reservoirs are also seen in our VISIR continuum data (see Sec. \ref{sec:MIR}), we are confident that the Br$\gamma$ gas emission areas are not remnants of the continuum subtraction. \cite{peissker2020b} show several dusty sources with a presumably stellar core at the position of the Br$\gamma$ features L0 and L1. Both features show an angular difference of $\Delta\,=\,60^{\circ}$. Compared to the 226 GHz radio map presented by \cite{Yusef-Zadeh2017} we find related counterparts to both L0 and L1. We speculate that also the additional gas and dust reservoirs associated with L2 - L4 (see Fig. \ref{fig:bar2005}) will contain compact dusty sources or young stellar objects. 

\subsection{Proper motion}
\label{sec:propermotion}
\begin{figure}[htbp!]
	\centering
	\includegraphics[width=.5\textwidth]{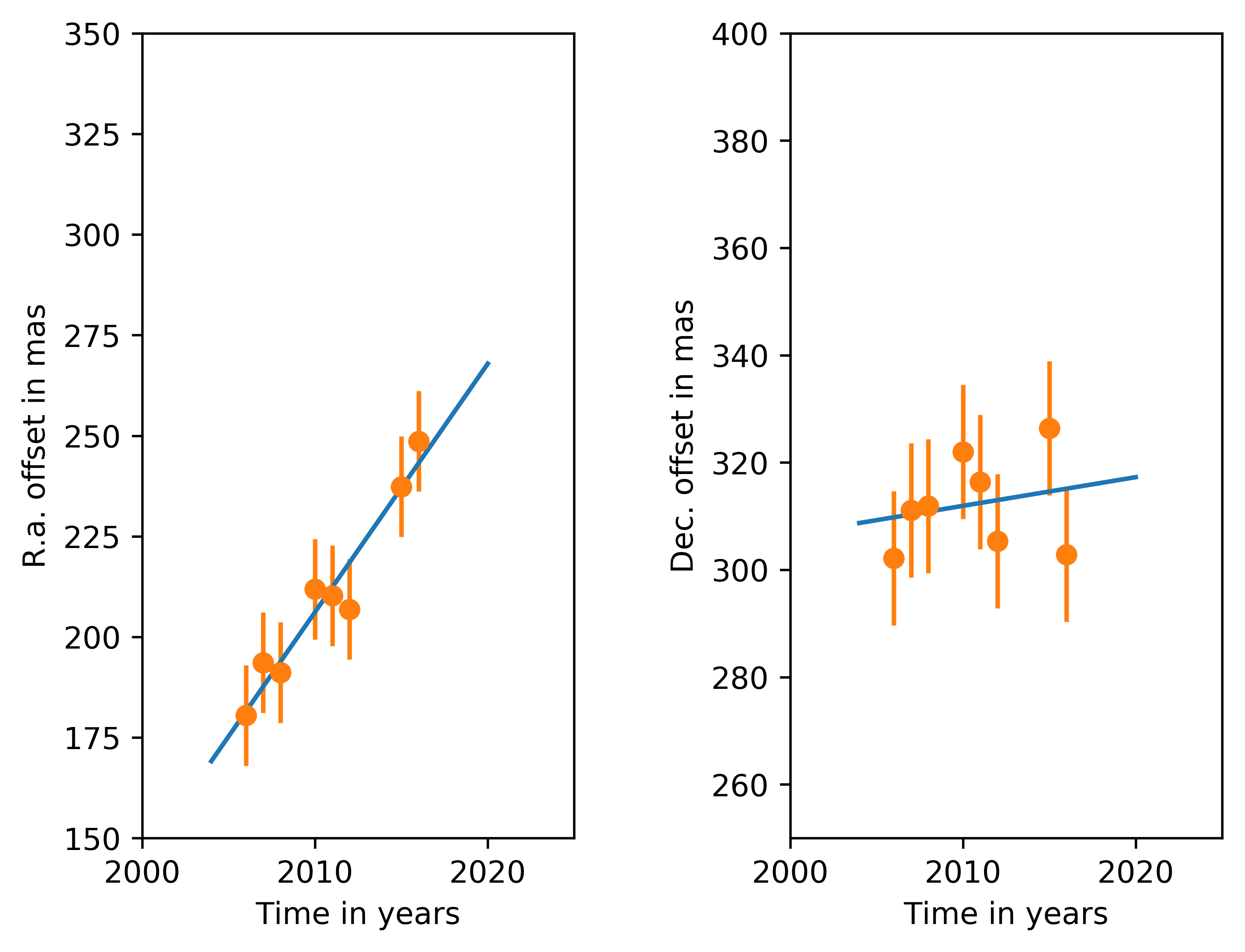}
	\caption{Proper motion of the prominent upper part of the Br$\gamma$-bar. The increasing distance of the bar towards Sgr~A* indicates a linear proper motion. The declination-offset is close to constant and justifies the limitation on the x-shifting vector for the cross-correlation method. From the fit, we are deriving a proper motion of $319\,\pm\,44$km/s.}
\label{fig:pm_plot}
\end{figure}

By comparing the line maps of the Br$\gamma$-bar, that are centered on Sgr~A*, a proper motion can be derived (Appendix, Fig. \ref{fig:pm}). As indicated by the contour maps shown in Fig. \ref{fig:bar}, we emphasize the analysis of the more prominent upper part of the Br$\gamma$-bar. To obtain the proper motion of this prominent part we use a cross-correlation method and additionally two Gaussian fitting methods (Fig. \ref{fig:pm_plot}).
For the former method, we shift the Br$\gamma$ line map of 2018 to the 2006 position of the Br$\gamma$-bar and record, at which shifting vector the emission reaches its maximum. The shifting vector is then equivalent to the distance, that the feature moved projected on the sky. We find a proper motion of $329\,\pm\,20$km/s directed westward.\newline %The lower part is showing two clumps with a distance of the maximum emission peak of around $0".12\,\pm\,0".01$. For these clumps, we find with the a distance to Sgr~A* a value of $86\,\pm\,20$km/s. We double checked the result by using a Gaussian, that is fitted to the lower part in 2006 and 2018. The distance is consistent with the cross-correlation method.
Since the Br$\gamma$ line flux distribution shows substructures (4 clumps, see Fig. \ref{fig:bar}), fitting a Gaussian to different substructures gives a complementary estimate of the velocities.
\begin{table}[hbt!]
    \centering
    \tabcolsep=0.08cm
    \begin{tabular}{ccccc}
         \hline %& flux [10$^{-16}$ erg/s/cm$^{2}$]
         \hline
           Year & R.A. & DEC.         &  $\Delta$R.A.     &  $\Delta$DEC.     \\% &  v$_{R.A.}$     &   v$_{DEC.}$     \\
                & in [mas] & in [mas] &   in [mas]    &  in [mas]             \\% &  in [km/s]     &   in [km/s]     \\
         \hline       
          2006  &   180.50   & 302.16 &   -    &    -     \\% &    -   &     -    \\
          2007  &   193.62   & 311.12 &  13.12   &  8.96  \\% &   512.48    &         \\
          2008  &   191.16   & 311.88 &  10.66   &  9.72  \\% &       &         \\
          2010  &   211.87   & 322.00 &  31.37   & 19.84  \\% &       &         \\
          2011  &   210.25   & 316.37 &  29.75   & 14.21  \\% &       &         \\
          2012  &   206.87   & 305.37 &  26.37   &  3.21  \\% &       &         \\
          2015  &   237.37   & 326.37 &  56.87   & 24.21  \\% &       &         \\
          2016  &   248.62   & 302.87 &  68.12   &  0.71  \\% &       &         \\
       \hline
    \end{tabular}
    \caption{Distance of the Br$\gamma$-bar. We measured the distance by using the substructures of the prominent upper part of the bar. The traveled distance $\Delta$R.A. and $\Delta$DEC. is related to the distance to 2006. The uncertainty is $\pm\,12.5$ mas and equivalent to $\pm\,1$ px.}
    \label{tab:pm_list}
\end{table}
\begin{table*}[hbt!]
    \centering
    \begin{tabular}{cccc}
         \hline %& flux [10$^{-16}$ erg/s/cm$^{2}$]
         \hline
           Method & v$_{R.A.}$ & v$_{DEC.}$        &  v$_{total}$ \\%& Uncertainty   &  Direction    \\% &  v$_{R.A.}$     &   v$_{DEC.}$     \\
                & [km/s] & [km/s] & [km/s]  \\%&  [km/s]      &     \\% &  in [km/s]     &   in [km/s]     \\
         \hline       
          Cross-correlation (upper part)  &  117   &  307        &   329   \\% & $\pm$ 20 & west   \\% &  -  &   -    \\
          Gaussian (upper part)  &   163   & 244      &  293    \\% & $\pm$ 60 & west  \\% &   512.48    &        \\
          Gaussian (lower part)  &   153  & 252       & 295     \\% & $\pm$ 16 & south  \\% &       &         \\
          Substructures (upper part)  &  209 & 241    &  319    \\% & $\pm$ 44 & west  \\% &       &         \\
          Substructures (lower part)  &   114   & 293 &  314    \\% & $\pm$ 15 & south  \\% &       &         \\
          \hline
    \end{tabular}
    \caption{Proper motion in R.A. and DEC. with different methods. For fitting a Gaussian to the upper and lower part, we use pixel values between 10-12. The clumpy substructures can be fitted with an 5-6 px Gaussian. Note that we shift the data-cubes of 2006 and 2018 for the cross-correlation only in R.A. direction. Typical uncertainties for $v_{total}$ are in the order of $\pm\,40$ km/s, for $v_{R.A.}$/$v_{DEC.}$ around $\pm\,20$ km/s.}
    \label{tab:pm_vel}
\end{table*}
From this, we get velocities for R.A. and DEC. that are based on the fit parameters (Fig. \ref{fig:pm_plot}). The resulting proper motion is $319\,\pm\,44$km/s directed westward.
For the sake of completeness, we are fitting a Gaussian to the prominent upper part in 2006 and 2018. We crop the lower part of the data-cube to exclude extra emission. This results in a proper motion of $293\,\pm\,60$km/s. The error includes different background scenarios, a variation of data quality, and emission, that might be suppressed due to stellar emission lines.

Combining all three velocities, we get $314\,\pm\,41$km/s for the prominent upper part of the Br$\gamma$-bar. 
Using a Gaussian fit to substructures and the complete lower part, we derive a total proper motion of $305\,\pm\,15$km/s. In contrast, the proper motion vector of the lower part of the Br$\gamma$-bar points to South whilst the upper part is directed towards West. Consult Table \ref{tab:pm_list} for the individual R.A. and DEC. proper motion velocities.
%For deriving the velocity, we use 1".0 as = 40 mpc and 1 pc = 3.08$\times\,10^{13}$km.
\subsection{Mid-infrared detection}
\label{sec:MIR}
\begin{figure*}[ht!]
	\centering
	\includegraphics[width=1.\textwidth]{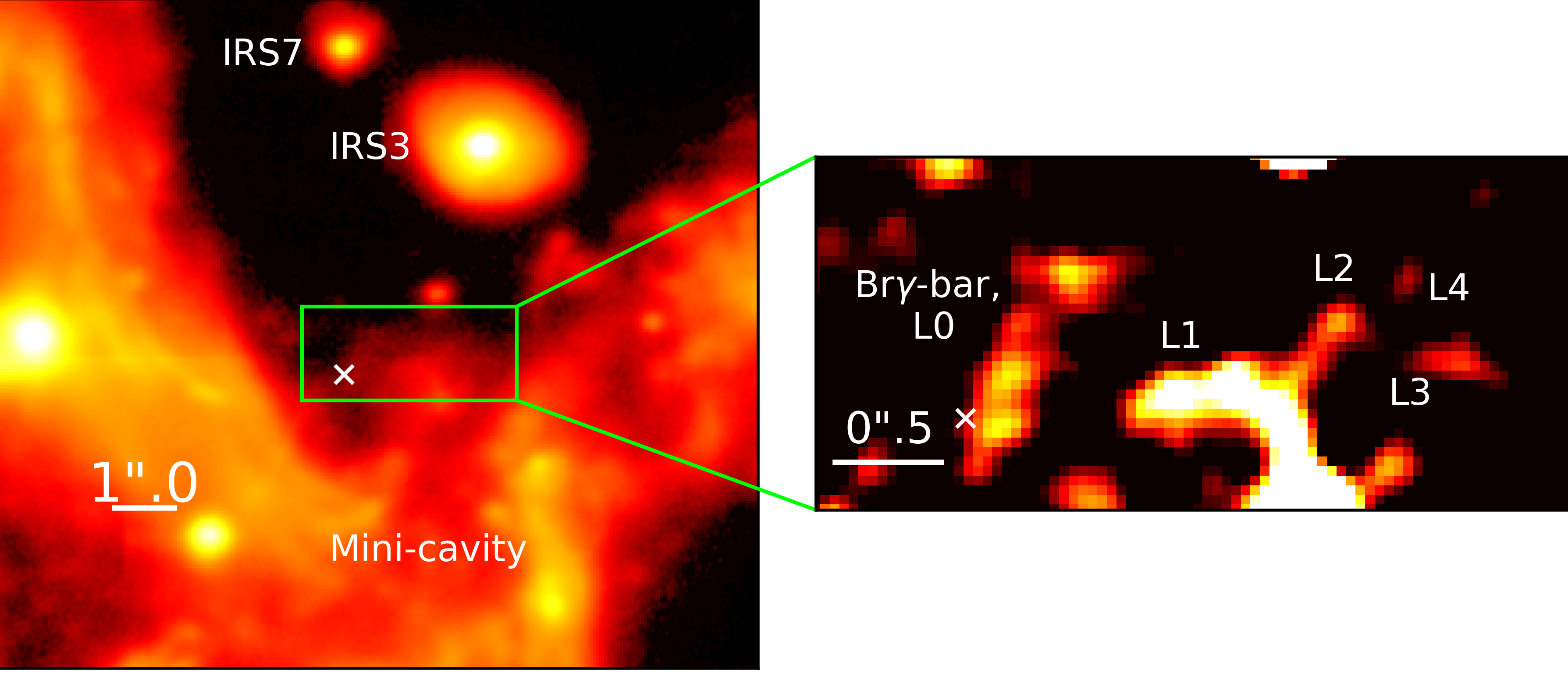}
	\caption{Overview of the GC in the MIR in 2016. The left image shows some prominent sources. The white cross marks the position of Sgr~A*. The right image is a low-pass filtered cutout of the green box displayed in the left overview map. For the name IDs, please see Fig. \ref{fig:bar2005}. The position of Sgr~A* was determined using the known positions of several SiO masers in the central cluster.}
\label{fig:mir_overview}
\end{figure*}

Based on the 2016 VISIR data set, we find at the position of the Br$\gamma$-bar in the N-band a PAH1 ($8.59\,\mu m$) counterpart with a width of $0.4\,\mu m$. The substructures shown in the contour plot in Fig. \ref{fig:bar} are indicated in the low-pass filtered cutout in Fig.\ref{fig:mir_overview}, right hand side.
The size of the Br$\gamma$ line flux distribution is in good agreement with the PAH1 detection. The angular difference between the 
L1 to L0 component is $\Delta\,=\,60^{\circ}$ and consistent with the Br$\gamma$-bar orientation (Fig. \ref{fig:bar2005}). Since the FOV of SINFONI is limited, we find a small size difference of around $10-15\,\%$ compared to the VISIR detection which is already included in the uncertainties of the analysis presented here. Whilst the size of the Br$\gamma$-bar in our SINFONI data can be determined to be around 0".95, the PAH1 continuum counterpart observed with VISIR extends over 1".11.  

\subsection{226 GHz radio detection}

The Br$\gamma$ features close to the line of sight towards Sgr~A* can 
also be identified in the radio. Here, we refer to the radio maps published by
\cite{Yusef-Zadeh2017-ALMAVLA} and in particular to their Fig.2 where the authors show several linear ridge components.
Unfortunately, the Br$\gamma$ features are very close to the Sgr~A*
position such that the corresponding identification can only be made
via elongations on the lowest contours of the Sgr~A* point source response.
\begin{table*}[htb!]
%\caption{}
\tabcolsep=0.1cm
\begin{center}
\begin{tabular}{cccccccccc}\hline \hline
Name ID &$\Delta$$\alpha$&$\Delta$$\delta$& description & S$_{226\,GHz}$&S$_{Br\gamma}$ - expected & S$_{Br\gamma}$ - observed & extinction coefficient\\
      & in [as]   & in [as]  & source size in [as] & in [mJy] & in [$10^{-17}\,Wm^{-2}$] & in [$10^{-17}\,Wm^{-2}$] & in [mag] \\  \hline
L0(EN/ES) &  0.0 to -1.50 & -0.3 to 0.4    & ridge &  $\sim$0.9   & $\sim$1.27 & 1.11$\,\pm\,0.3$ & $\sim$2.84\\
      &               &                & $>0.8\,\times\,<0.2$ &   &  & & \\
L1    & -0.3 to -0.70 & -0.14 to 0.06  & ridge &1.3$\,\pm\,$0.4 & 1.84$\,\pm\,$0.6 & 1.54$\,\pm\,$0.6 & 2.97$\,\pm\,$0.7 \\
      &               &                & 0.4$\times <0.2$ & & &  & \\
L2    & -0.8 to -1.40 &  0.04 to 0.25  & 3 sources&1.3$\,\pm\,$0.4 & 1.84$\,\pm\,$0.6 & 1.55$\,\pm\,$0.4$^*$ & 2.94$\,\pm\,$0.6$^*$ \\
      &               &                & all $<0.2$ &         & & & \\
L3    & -1.5 to -1.70 &  0.00 to 0.20  & $<0.2$ &1.2$\,\pm\,$0.3 & 1.70$\,\pm\,$0.4 & 1.31$\,\pm\,$0.4 & 3.26$\,\pm\,$0.8\\
L4    & -1.8 to -2.00 &  0.10 to 0.30  & $<0.2$ & 0.6$\,\pm\,$0.2 & 0.85$\,\pm\,$0.3 & 0.83$\,\pm\,$0.4 & 2.54$\,\pm\,$0.7\\
\hline \hline
\end{tabular}
\end{center}
\caption{Linear ridge components as outlined in Fig. \ref{fig:bar2005}. We list relative positions to Sgr~A* and extent of the aperture within which most of the flux is contained, description of the source component and a limit on the size, 226 GHz flux densities based on the observations by \cite{Yusef-Zadeh2017} and expected Br$\gamma$ line fluxes calculated with Eq. \ref{eq:expected_flux}. The observed flux is based on the SINFONI H+K data cubes. The last column shows the derived extinction coefficient. The identification of the sources west of Sgr~A* is guided by the shape of the radio contours. Size estimates and distances are given in arcseconds. Note: the ($^*$) marks 3 single detections that are, based on the radio observation, grouped together as L2. The errors of the expected Br$\gamma$ flux is pervaded from the observed radio detection. 
The radio flux for the L0 component is not known. 
Here, it is assumed to be on the same order as the 
flux densities of the other components. The uncertainties of the observed Br$\gamma$ flux is based on the SINFONI flux analysis.}
\label{tab:names}
\end{table*}

In Fig. \ref{fig:bar2005} we compare the smoothed Br$\gamma$ map with the radio map.
In Table \ref{tab:names} we list the components names, the relative coordinates and their 226 GHz fluxes.
We identify the northern and southern radio extensions marked EN and ES
with the northern and southern tip of the north-south (NS) Br$\gamma$ feature.
The western Br$\gamma$ feature can be identified as the extension labeled $L1$.
It is followed by the curved source complex labeled $L2$ and the extended 
(0.15 to 0.2 arcseconds diameter) sources $L3$ and $L4$.
The 226 GHz radio flux density of these components can be estimated
from their extent and the contour line labels given in Fig.2 by 
\cite{Yusef-Zadeh2017-ALMAVLA}.

The east-west linear feature L1 appears to arise from Sgr~A*.
\cite{Yusef-Zadeh2017-ALMAVLA} points out that it coincides with the 
diffuse X-ray emission and a minimum in the near-IR extinction.
They argue that the millimeter emission is produced by synchrotron 
emission from relativistic electrons in equipartition with 
an $\sim$1.5 mG magnetic field. 

\cite{Yusef-Zadeh2017-ALMAVLA} argue that
the linear ridge may become brighter close to the peak of
Sgr~A* at 226 GHz and may therefore have 
a hard or highly inverted spectrum since there is no detectable 
emission of it at lower frequencies. 
However, at lower frequencies the beams typically become larger and 
sources close to Sgr~A* are more difficult to be detected. Hence,
the scenario could also be compatible with a flat spectrum 
Bremsstrahlung source.

\cite{Yusef-Zadeh2017-ALMAVLA} point out that in the recent milli-arcsec resolution, the 86 GHz and 230 GHz observations of Sgr~A* show indications of an asymmetric source structure \citep{Brinkerink2016, Fish2016}.
Towards the east under a position angle of about 90 degrees, a secondary source shifted by 100 micro-arcsec from Sgr~A*
is indicated by the data. While this could be explained by interstellar scattering \citep{Brinkerink2016}, intrinsic source properties could be responsible as well.

However, given the proper motion of the ridge component L0 in the order
of 320 km/s, it is very likely that it is part of the minispiral gas flow. Since L1 is comparable in shape and size to L0, it is very likely that the east-west feature could also be part of this gas flow.

Assuming an optically thin free-free emission spectral index
of 0.1 between 5 GHz and 226 GHz and an intrinsic 
Br$\gamma$/Br$\alpha$ ratio of 2.8, we can estimate the expected Br$\gamma$ line intensity
from the 226 GHz radio continuum flux density using the
formula \citep{Glass1999}\newline
%\begin{equation}
%I(Br\gamma)=0.97 \times 10^{-14}(\frac{T}{10^4 K})^{-0.85}
%(\frac{\nu}{5 GHz})^{0.1} F_{\nu}(Jy)\,\,\,Wm^{-2}
%\label{eq:expected_flux}
%\end{equation}
\begin{equation}\label{eq:expected_flux}
\begin{split}
& I({\rm Br}\gamma) = \\
& 1.42  \times 10^{-17}\left(\frac{T}{10^4\,{\rm K}}\right)^{-0.85}\left(\frac{\nu}{226\, {\rm GHz}}\right)^{0.1} F_{\nu}({\rm mJy})~{\rm Wm^{-2}}
\end{split}
\end{equation}

%with $I(Br\gamma) = 1.27\,\times\,10^{-17}\,Wm^{-2}$. Furthermore, we observe a flux density of $I(Br\gamma)_{obs} = 1.11\,\pm\,0.3\,\times\,10^{-17}\,Wm^{-2}$ in the SINFONI Br$\gamma$ line map. It is related to the channels 1515-1520 that corresponds to $2.16527\,\mu m$ and $2.16825\,\mu m$.
%From the ratio $x$ of the expected to observed flux density, we obtained an estimate of the extinction with $2.5^x$ in the mini-spiral region at $2.16\,\mu m$. 
We then compare the predicted flux density $I(Br\gamma)$ to the observed flux density that is based on the SINFONI Br$\gamma$ line map detection. These maps are related to the channels 1515-1520 that corresponds to $2.16527\,\mu m$ and $2.16825\,\mu m$. From the ratio of the expected to observed flux density, we obtained an estimate of the extinction in the mini-spiral region at $2.16\,\mu m$. The results are listed in Table 2.

The observed extinction ranges from 2.5-3.2 mag, with an averaged extinction of $2.91\,\pm\,0.72$ mag (see Table \ref{tab:names}). This is in good agreement with \cite{Schoedel2010}, who reported a median extinction value of $A_{K_s}=2.74\,\pm\,0.30$ mag in the range of 1.8-3.8 mag in the mini-spiral arms, using H-K colors. 

\section{Discussion}
\label{sec:discussion}
In this section we will discuss the results of our analysis. We also link the observed filaments to large scale features in the region.

\subsection{Distance estimate of Br$\gamma$ bar and its stability in the hot bubble}

Based on the tangential velocity of $v_{\rm t}\sim 314\,{\rm km\,s^{-1}}$ and the radial velocity of the bar $v_{\rm r}\sim 53\,{\rm km\,s^{-1}}$ (based on the Br$\gamma$ emission line), we estimate the total space velocity using $v_{\rm bar}\sim (v_{\rm t}^2+v_{\rm r}^2)^{1/2}\sim 318\,{\rm km\,s^{-1}}$. From this, we may approximately calculate the distance from Sgr~A* assuming a bound circular orbit, $r_{\rm bar}\sim GM_{\bullet}/v_{\rm bar}^2\sim 0.17\,(M_{\bullet}/4\times 10^6\,M_{\odot})(v_{\rm bar}/318\,{\rm km\,s^{-1}})^{-2}\,{\rm pc}\sim 4.25''$. In case the Br$\gamma$ is unbound and moving along the parabolic orbit, the distance would be a factor of two larger, $r_{\rm bar}^{\rm esc}\approx 2GM_{\bullet}/v_{\rm bar}^2\sim 0.34\,{\rm pc}=8.5''$. In the following discussion, we assume the bar is bound and hence the smaller distance estimate applies.

The associated orbital (dynamical) timescale is $t_{\rm dyn}\sim 2\pi [r^3/(GM_{\bullet})]^{1/2}\sim 3300$ years. This applies for the more prominent westward-moving upper part of the Br$\gamma$ bar. Additionally, we derive similar values for the southward-moving lower part of the bar, specifically we determine $r_{\rm bar}\sim 0.18\,{\rm pc}$ and $t_{\rm dyn}\sim 3600$ years.

With $r_{\rm bar}\sim 0.17\,{\rm pc}$, the Br$\gamma$ bar is located at or just beyond the Bondi radius, $r_{\rm Bondi}\sim GM_{\bullet}/c_{\rm s}^2$, where $c_{\rm s}$ is the sound speed of the hot gas that is fueled mostly by OB stars. This hot plasma has the temperature of $T_{\rm Bondi}\sim 10^7\,{\rm K}$ and the number density of $n_{\rm Bondi}\sim 26\,{\rm cm^{-3}}$, from which $r_{\rm Bondi}\sim 4''(T_{\rm Bondi}/10^7\,{\rm K})\sim 0.16\,{\rm pc}$ \citep{2003ApJ...591..891B,2013Sci...341..981W}. In terms of stellar populations, the Br$\gamma$ bar can be found at larger scales than the S-cluster ($1''\sim 0.04\,{\rm pc}$), potentially within the Clockwise Disk (CWD) of young massive OB stars located between $0.032-0.48\,{\rm pc}$ \citep{2009ApJ...697.1741B}. The derived distance is also comparable to the pericentre distances of the Northern and the Eastern Arm streamers of the minispiral \citep{Zhao2009}, $q_{\rm NA} \simeq 0.17\,{\rm pc}$ and $q_{\rm EA} \simeq 0.25\,{\rm pc}$, respectively, which raises the possibility that the Br$\gamma$ bar is a detached minispiral material that appears to be positioned close to Sgr~A* just in projection. The Br$\gamma$ bar could also have originated in the collision site of the Northern and the Eastern arm, so-called the Bar, located at $\sim 0.1-0.2\,{\rm pc}$ south of and behind Sgr~A* \citep{2010ApJ...723.1097Z}.  

When one compares the gas pressure of the bar, $P_{\rm bar}=n_{\rm bar}k_{\rm B}T_{\rm bar}\sim 2.6\times 10^5\,{\rm cm^{-3}}\times 1.38\times 10^{-16}\,{\rm erg\,K^{-1}}\times 10^4K\sim 3.588\times 10^{-7}\,{\rm erg\,cm^{-3}}$, with the pressure of the bremsstrahlung plasma at the Bondi radius, $P_{\rm Bondi}=n_{\rm Bondi}k_{\rm B}T_{\rm Bondi}=26\,{\rm cm^{-3}}\times 1.38\times 10^{-16}\,{\rm erg\,K^{-1}}\times 10^7\,{\rm K}\sim 3.588\times 10^{-8}\,{\rm erg\,cm^{-3}}$, we see that the bar pressure is larger than the pressure at the Bondi radius by one order of magnitude. This implies that the bar is not confined by the surrounding thermal pressure. In that case, the bar expands with a velocity approximately equal to its sound velocity, $v_{\rm exp} \sim c_{\rm bar}=[k_{\rm B}T_{\rm bar}/(\mu m_{\rm H})]^{1/2}\sim 13\,{\rm km\,s^{-1}}$.\\ 
Due to the relative motion of the bar and the ambient medium fueled by stellar winds, the force arises on the bar because of to the ram pressure, $P_{\rm ram}\sim \rho_{\rm Bondi}v_{\rm rel}^2$, where the relative velocity is the vector sum of the bar motion with respect to the stellar wind velocity field. Given $v_{\rm bar}\sim 318\,{\rm km\,s^{-1}}$ and the average stellar-wind velocity of $v_{\rm wind}=750\,{\rm km\,s^{-1}}$ \citep{Najarro1994}, we set $v_{\rm rel}\sim 10^3\,{\rm km\,s^{-1}}$. The ram pressure then is $P_{\rm ram}=\mu m_{\rm H} n_{\rm Bondi} v_{\rm rel}^2=0.5\times 1.67\times 10^{24}\,{\rm g}\times 26\,{\rm cm^{-3}}\times (10^8\,{\rm cm\,s^{-1}})^2\sim 2.2\times 10^{-7}\,{\rm erg\,s^{-1}}$. The ram pressure is of a comparable order of magnitude as the thermal pressure of the Br$\gamma$ bar, and hence can confine the bar externally or at least influence its further evolution. 

Moreover, dynamically important magnetic field in the vicinity of Sgr~A* can affect the evolution of Br$\gamma$ bar more than the combined effect of the ram pressure and the external thermal pressure. The magnetic pressure can be estimated from the Faraday rotation measurements of the magnetar PSR J1745-2900, which is located at the comparable distance as the Br$\gamma$-bar, $r\gtrsim 0.12\,{\rm pc}$ from Sgr~A*. The Faraday rotation measurements indicate a line-of-sight component of the magnetic field of $B\gtrsim 8\,{\rm mG}$ \citep{2013Natur.501..391E}, from which the magnetic pressure can be estimated as $P_{\rm mag}=B^2/8\pi\approx 2.6\times 10^{-6}\,{\rm erg\,cm^{-3}}$ and we see that $P_{\rm mag}>P_{\rm bar}$. It is now premature to say whether the Br$\gamma$-bar belongs to the filamentary structures such as nonthermal filaments that follow the global magnetic field lines. \cite{2017ApJ...850L..23M} reported about the thin filamentary structure to the north and the east of Sgr~A*, so-called Sgr A West Filament (SgrAWF). In their radio images at centimeter wavelengths as well as Pa$\alpha$ image, one can also see the potential counterparts of the L0-L4 complex to the west of Sgr~A*. We hypothesize at the moment that the Br$\gamma$ bar reported here could be the manisfestation of the minispiral material which is shaped and dynamically affected by the global poloidal magnetic field in a similar way as SgrAWF or by the external magnetohydrodynamic drag. However, more multiwavelength data is necessary to make definite conclusions. We discuss more mechanisms responsible for clumpy filamentary structures in the Galactic center region in the following subsections.   

The bar as a whole will inevitably tend to be tidally stretched since its number density expressed by Eq.~\ref{eq_bar_clump_density} is five orders of magnitude smaller than the Roche condition of the self-gravitational collapse derived for fully ionized plasma at the estimated distance of the bar,
\begin{equation}
\begin{split}
    & n_{\rm bar}\ll n_{\rm Roche}\,=\, \\ 
    & 4.73\times 10^{10} \left(\frac{M_{\bullet}}{4\times  10^6\,M_{\odot}}\right) \left(\frac{r}{0.17\,{\rm pc}}\right)^{-3}\,{\rm cm^{-3}}\,.
    \label{eq_roche}
\end{split}
\end{equation}
In addition, the tidal radius $r_{\rm t}$ for the bar half-length of $R_{\rm bar}\sim 0.5''\sim 0.02\,{\rm pc}$ (based on SINFONI and VISIR), the estimated bar mass of $m_{\rm bar}\approx 2.5\times 10^{-5}\,M_{\odot}$, and the Sgr~A* mass of $M_{\bullet}=4\times 10^6\,M_{\odot}$ is
\begin{align}
    r_{\rm t} & = R_{\rm bar}\left(\frac{2M_{\bullet}}{m_{\rm bar}}\right)^{1/3}=\notag \\
              & = 137\left(\frac{R_{\rm bar}}{0.5''}\right)\left(\frac{m_{\rm bar}}{2.5 \times 10^{-5}\,M_{\odot}} \right)^{-1/3}\,{\rm pc}\,,
    \label{eq_tidal_radius}          
\end{align}
hence the bar is prone to the complete tidal disruption since it moves around Sgr~A* at $r\approx 0.2\,{\rm pc}$, which is three orders of magnitude closer than $r_{\rm t}$. One still cannot exclude the possibility that Br$\gamma$ bar is a filamentary, clumpy structure, which is being overall tidally disrupted, but individual clumps would be self-gravitating with number densities of the order of the Roche critical density, see Eq.~\eqref{eq_roche}. These clumps would, however, have to be very compact with length-scales of a few {\rm AU}, which follows from the Hill radius,
\begin{align}
r_{\rm Hill} & =r_{\rm bar}\left(\frac{m_{\rm cl}}{3M_{\bullet}}\right)^{1/3}= \notag \\
             & = 3.13\left(\frac{r_{\rm bar}}{0.17\,{\rm pc}}\right)\left(\frac{m_{\rm cl}}{2.1\,M_{\oplus}}\right)^{1/3}\,{\rm AU}.
  \label{eq_hill_radius}           
\end{align}
Individual clums could also hide stars that would be enshrouded in optically-thick gaseous-dusty shells with the length-scale of $r_{\rm Hill}\sim 153\,{\rm AU}$ \citep{Zajacek2014,Valencia-S.2015,Shahzamanian2016,Zajacek2017}, which follows directly from Eq.~\eqref{eq_hill_radius} for $m_{\rm cl}=1\,M_{\odot}$. However, any association of the Br$\gamma$ bar with the population of dust-enshrouded objects \citep{peissker2020b,Ciurlo2020} is currently speculative and requires additional monitoring of the dynamics of this Br$\gamma$ complex. 

For estimating the lifetime and the stability of the Br$\gamma$ bar, we follow \citet{Burkert2012}, who derived basic timescales for the interaction of a colder clump, in our case a filament or a streamer, with the surrounding hot medium. The ablation of material due to the motion through the hot medium is rather slow with the ablation timescale of
\begin{align}
    \tau_{\rm abl} & =\frac{R_{\rm bar}}{0.25 q_{\rm abl}v_{\rm bar}}\frac{\rho_{\rm bar}}{\rho_{\rm Bondi}}=6\times 10^8\left(\frac{R_{\rm bar}}{0.02\,{\rm pc}}\right)\times \notag \\
    &\times \left(\frac{v_{\rm bar}}{318\,{\rm km\,s^{-1}}}\right)^{-1} \left(\frac{\rho_{\rm bar}/\rho_{\rm Bondi}}{10000}\right)\,{\rm yr}\,,
\end{align}
where we considered $q_{\rm abl}\approx0.004$. The evaporation due to thermal conduction proceeds much faster. In the saturation limit and assuming pressure equilibrium (that may not apply as we estimated earlier), we obtain
\begin{equation}
    \tau_{\rm evap}\approx 177\,\left(\frac{r_{\rm bar}}{0.17\,{\rm pc}} \right)^{1/6}\left(\frac{m_{\rm bar}}{2.5\times 10^{-5}\,M_{\odot}} \right)^{1/3}\,{\rm yr}\,, 
\end{equation}
which is an order of magnitude less than the dynamical timescale of a few thousand of years. This implies that the Br$\gamma$ bar is likely a temporary feature that will not survive long enough to affect the activity of Sgr~A*, since the free-fall timescale from $r_{\rm bar}$ is,
\begin{equation}
    t_{\rm ff}=\pi\left(\frac{r_{\rm bar}^3}{GM_{\bullet}}\right)^{1/2}\sim 1643\,{\rm yr}\,.
    \label{eq_free_fall_timescale}
\end{equation}
Due to the velocity shear between the bar and the Bondi plasma, the bar is susceptible to the Kelvin-Helmholtz (KH) instability. The velocity shear can be assumed to correspond to the orbital velocity of the bar, $v_{\rm shear}\sim v_{\rm bar}\sim 318\,{\rm km\,s^{-1}}$. Given the density ratio of $r=n_{\rm Bondi}/n_{\rm bar}\sim 10^{-4}$ between the hot plasma and the bar, the instabilities of the size comparable to individual clumps, $\lambda_{\rm cl}\sim 3\,{\rm mpc}$, will develop on the timescale of
\begin{equation}
    \tau_{\rm KH}=\frac{\lambda_{\rm cl}}{v_{\rm shear}}\frac{1+r}{\sqrt{r}}\approx 923\,{\rm years}\,.
    \label{eq_kh_timescale}
\end{equation}
Hence, the KH instabilities with the size comparable to the observed clumps can develop within one orbital (free-fall) timescale and the observed uneven surface brightness of the Br$\gamma$ structure can already be a manifestation of this process. The magnetic field can in theory suppress the formation of Kelvin-Helmholtz instabilities and stabilize the structure, especially if the internal tangled magnetic field is present in the bar structure itself \citep{2015MNRAS.449....2M}. On the other hand, as found in \citet{2015MNRAS.449....2M}, the structure would still fragment into clumps that would not mix into the hot medium but rather co-move with it. Hence, the bar would continue to change its appearance and the surface brightness during the timescales derived above.  

In summary, the Br$\gamma$ bar is a temporary feature that will likely disappear within 100--1000 years due to the combined action of evaporation, tidal stretching, and instability development. Given its estimated distance of $r_{\rm bar}\sim 0.2\,{\rm pc}$ from Sgr~A*, it is unlikely to affect its activity since the free-fall timescale is longer than both the evaporation and the instability timescale. These conclusions could in principle be affected by a complicated interplay of different components in the central parsec (minispiral, stars, Sgr~A*, dust). For instance, if the bar was formed due to the stellar wind-wind collisions as modelled by \citet{Calderon2016,Calderon2020}, the structure can be continuously recreated and/or formed elsewhere depending on the spatial distribution of stars inside the nuclear star cluster. However, \citet{2020MNRAS.493..447C} analyzed the cold clump formation via the non-linear thin shell instability in the wind-wind interaction and found a small mass of individial clumps in the range of $\sim 10^{-3}-10^{-2}\,M_{\oplus}$, which is much smaller than the mass of a few Earth masses inferred for the clumps along the bar. This makes the association of the Br$\gamma$ bar with the minispiral streamers more likely since the Northern and Eastern streamers alone contain $\sim 12\,M_{\odot}$ of ionized gas \citep{2010ApJ...723.1097Z}. We will discuss the relation of the bar to other structures in the GC region in the following subsections.

\subsection{Thin filaments of the GC}
As shown by \cite{muzic2007}, elongated and narrow filaments in the vicinity of Sgr~A* seem to be the rule and not the exception \citep[see also][]{Ciurlo2014, Ciurlo2019}.

Even on larger scales compared to the observations presented in  \cite{muzic2007}, filaments with a specific orientation with respect to the central region associated with Sgr~A* can be observed. Examples are presented, e.g. in the inner $1^{\circ}\times 1^{\circ}$ by \cite{LaRosa2004}. Here, the authors characterize their detection as non-thermal filament (NTF) candidates (for larger scales see also \cite{YusefZadeh2005, Morris2017}).
Since the  Br$\gamma$ feature we investigate here, is presumably influenced by the UV radiation of hot massive young He-stars \citep{YusefZadeh1996, Carlsten2018, Kim2018} at the center of the GC stellar cluster potentially including the S-stars, the gas can very likely considered to be a thermal filament (TF) dominated by Bremsstrahlung emission (see Sec. \ref{sec:stellarwind}). However, the NTF candidates morphology is shaped by the magnetic fields of the GC. These magnetic fields imply a non-poloidal shape which results furthermore in a complex structure. In order to differentiate between the TF and NTF character of the filaments, polarization measurements of the features at high angular resolutions are needed.

%In the upcoming Sec. \ref{sec:meissner}, we will discuss the possibility that the here analyzed thermal Br$\gamma$-bar is aligned because of magnetic field lines \citep[see][]{Zhang2017, Rigoselli2019}. The origin of these field lines is assumed to be created by a magnetic filed created by Sgr~A*.

As shown in Sec. \ref{sec:propermotion}, the proper motion of the upper part of the Br$\gamma$-bar is around 320 km/s and therefore comparable with the proper motion of other filaments detected in the infrared in the vicinity of Sgr~A*. For example,
following the  nomenclature by \cite{muzic2007}, NE3, NE4, and SW7 are showing proper motions that are in the same range.

\subsection{Location of the Br$\gamma$ bar features}

While the Br$\gamma$ bar features we describe here are 
located close to the line of sight towards Sgr~A*, one can raise the question of what their true physical distance to the SMBH and the S-star cluster is.
Very close proximity to Sgr~A* on scales of the S-cluster members would
imply high orbital velocities of several hundred to a few thousand 
km/s. Also the gaseous features would likely be distorted rather than
almost linear. Hence, it is unlikely that they are in the immediate vicinity 
of Sgr~A*. However, they could be part of the general mini-spiral
flow. It seems peculiar that they are located in projection north 
of Sgr~A* rather than south were the bulk of the material is passing by Sgr~A*.
However, ALMA observations show that in the central few arcsecond and 
just north of Sgr~A* complex, extended emission can be found in the emission
of density tracers like CS and H$_3$CO$^+$ \citep{Moser2017}.
In fact, kinematic modeling of the gaseous mini-spiral material indicates that 
the gas is rather turbulent.

\cite{VollmerDuschl1999} present a
self-consistent hydrodynamic
model of the gas flow in the mini-spiral system.
They model the mini-spiral gas streams as three disk
systems that are interleaved and form structures like the
northern arm and the mini-spiral bar feature.
In their modeling, the authors relate to the
observations of the 
Sgr A West complex in the H92$\alpha$ line
at 8.3 GHz with a resolution of 1 arcsec as 
presented by \cite{RobertsGoss1993}
as well as the [Ne II]($\lambda\,12.8\mu$m) line
emission observations described in
detail by \citep{Lacy1991}.
Similar models of the mini-spiral gas have also been
put forward by e.g.  \cite{Zhao2009} or \cite{Tsuboi2017}.

In order to match the observations, 
\cite{VollmerDuschl1999} allow for
an ad-hoc radial accretion velocity of
5\% of the local Keplerian azimuthal
velocity.
They also find, that a satisfactory 
description of the data is only possible if
they allow for a 
disk-like flow with a turbulent velocity 
of 40\% of the Keplerian velocity,
and a vertical turbulent velocity of 5\% of the
Keplerian velocity.
The turbulent velocities determine the thickness and viscosity of the disk.

The authors point out that in the framework of the standard
theory of accretion disks
\citep[see, e.g][]{Frank1992}, this
corresponds to a viscosity parameter
$\alpha\,\sim\,0.3$. This is a value which is 
well within the range found for 
other types of disks.
In Fig.17 of \cite{VollmerDuschl1999}, the authors plot the modeled disk material
at low intensity levels and find that a tenuous gas
component may fill almost the entire planes that 
represent gaseous stream of the northern arm and the bar.

Therefore, it seems conceivable that the Br$\gamma$-bar features 
we report are part of the turbulent mini-spiral as a stream that 
passes by Sgr~A* from behind (see geometry of the model presented by
\cite{VollmerDuschl1999}). This is consistent with the fact that
no strong temporal variations in extinction, i.e. reddening, towards the
S-cluster members or even the infrared counterpart of Sgr~A* have 
been reported.

\subsection{Stellar wind induced ionization}
\label{sec:stellarwind}

Based on the detection of the bar throughout the available SINFONI data, the emission of the ionized feature in the vicinity of Sgr~A* is the focal point 
of several phenomena in the GC. The case B recombination lines imply UV radiation by the surrounding stars \citep{Najarro1994, Najarro1997, Martins2007}. The authors of \cite{Karas2019} model the combination of a wind originating in a stellar cluster and the ISM and describe two shocks: a wind stellar wind shock and a jet-shock induced by the ISM. The transition zone between the wind- and jet-shock, that harbors clumps, is discussed in the following subsection. 
\cite{Calderon2016, Calderon2020} modeled and discussed the influence of the stellar winds of the Stars in the GC on scales of a few parsecs. \cite{Ressler2018} shows even closer simulations of the inner parsec. The massive and luminous He-stars are the major wind sources. The lower luminous S-cluster stars may contribute to the wind impact onto the filaments only in the case where the Br$\gamma$ filaments we describe here are close to the central arcsecond.
The authors describe that the S-star S2 and complex structures in the accretion wind have an influence on the low accretion rate that they derive to be a few $\times\,10^{-8}\,M_{\odot}\,yr^{-1}$. 

\subsection{Shock induced clumpiness}
\label{sec:shock_induced_clumps_h2}
As shown in Fig. \ref{fig:bar}, the spectrum of the Br$\gamma$-bar shows a prominent [FeII] and H$_2$ line. Both emissions are indicators for shocks induced by stellar winds of young OB stars or an accretion wind onto and from Sgr~A*. As already mentioned in Sec. \ref{sec:spectral_emission_lines}, the H$_2$S(0) line seems to be connected to a different object because of the large blue-shifted line-of-sight velocity (see discussion in Sec. \ref{sec:H2_line}). %We speculate, that the location of this stellar object could be embedded in the Br$\gamma$-bar. 

%Since Fig. \ref{fig:bar} is the co-added result of the SINFONI data between 2005 and 2015, the object that is connected to the Doppler shifted H$_2$S(0) line could be an S-star or a dusty source \citep{Ciurlo2020, peissker2020b}.
However, the [FeII] emission can be associated with fast collisions of winds and gas clouds and as a consequence, shocks could be the reason for the ionization of the gas \citep{Mouri2000}. These shocks could then also be responsible for the observed and detected clumps as already simulated and discussed by \cite{Calderon2016}. Our estimated lower mass limit of $2.5\,\times\,10^{-5}\,M_{\odot}$ is consistent with the mass, that the authors derive for the clumps that are created by the collision of stellar winds produced by the S-stars. 

\subsection{The blue-shifted H$_2$S(0) line}
\label{sec:H2_line}

We find, that the L0 Br$\gamma$-bar feature is located at the position of the gap seen in the ALMA detection of a rotating disk structure \citep{Murchikova2019}. The authors report, that the gap separates the red- and blue-shifted part of the disk spectrum. Within the uncertainties and the width of the rotating ALMA disk spectrum, our Br$\gamma$-bar feature L0 may be responsible for an absorption resulting in the detected gap \citep[see corresponding comment in][]{Murchikova2019}. This would make the bar feature a foreground object with respect to the structure detected with ALMA.
Alternatively, the Br$\gamma$-bar L0 could be located at the center of the rotating ALMA disk, absorbing only the background disk emission. If the disk is orbiting Sgr~A* this would position the bar feature close to the center. This, however, would imply a high velocity spread (several 100 km/s, similar to the S-star cluster members) of the gas constituting the bar feature. This is not observed, hence, making this explanation a more challenging and a though less likely one.

%This is also reflected by the H$_2$/Br$\gamma$ line ratio that we derive to be less than 0.4.

% \subsection{Accretion of Sgr~A*}
% 
% According to \cite{Quataert2000}, the accretion rate of Sgr~A* is just a fraction of a solar mass M$_{\odot}$. The %authors  derive a density of the surrounding material of Sgr~A* to about $\sim\,10^6\,cm^{-3}$ that results in a %accretion of about  $10^{-8}\,M_{\odot}\,yr^{-1}$. We assume, that the here described Br$\gamma$-bar is most likely %created by stellar winds  and shocks. From that we speculate, that the stellar winds, that contribute to the low %accretion rate of Sgr~A*, also  create the ionized emission of the Br$\gamma$-bar. Recently, \cite{Ressler2018} derived a %accretion rate for Sgr~A* a few  $\times\,10^{-8}\,M_{\odot}\,yr^{-1}$ which is consistent with value provided by %\cite{Quataert2000}. The authors of  \cite{Ressler2018} use $\approx$30 Wolf-Rayet stars to simulate multiple complex %stellar wind interactions. We see an  argumentative overlap of the here presented observed and already described %phenomena in e.g. \cite{Najarro1994,  Quataert2000, Martins2007, Yusef-Zadeh2012, Calderon2020}. The constant shape in %combination with the derived proper  motion of the Br$\gamma$ feature indicates, that the elongated gas reservoir is not %affected by the low accretion rate of  Sgr~A*.

\subsection{Upcoming MIRI observations}

Considering the mid-infrared detection of features close to Sgr~A*, the upcoming MIRI instrument on board of the James-Webb-Space-Telescope (JWST) \citep[consider][]{Bouchet_2015, Rieke_2015, Ressler_2015} is a reasonable opportunity for observations. Even though the plate scale of MIRI is slightly increased compared to SINFONI, the size of the observed features ensure a continuum detection with MIRI. This is even more true considering the comparable plate scale of VISIR and NIRSpec and the detection of the Br$\gamma$-bar in the MIR (see Fig. \ref{fig:mir_overview}). With NIRSpec we will have also access to a confusion free Doppler shifted Pa$\alpha$ emission line because of the not present telluric absorption and emission features. We are expecting a more complete picture of the observed structures. With the spectroscopic capabilities of MIRI, we will have broad band information about several emission lines like e.g., $CO$, $SiO$, and $CH_4$ \citep[][]{peissker2020b}. The stable point spread function of the JWST will also help to image
faint and extended features in the central stellar cluster.

\section{Conclusions}
\label{sec:conclusions}

In this work, we have shown that the Br$\gamma$-bar located to the west of Sgr~A* is in projection close to the S-cluster and exhibits a proper motion of around 320 km/s in westward direction. This velocity matches proper motion values of other infrared and radio features traced in that region \citep{Yusef-Zadeh2004, muzic2007, Yusef-Zadeh2017-ALMAVLA}. Altough it is close to Sgr~A* in the line of sight, we argue that it is at least part of the mini-spiral passing by the very center. We find substructures that indicate a clumpiness that are most plausibly caused by the shocks of stellar winds. The lower limit for the mass of the observed Br$\gamma$-bar feature in the SINFONI FOV is $2.5\,\times\,10^{-5}\,M_{\odot}$. This value is in agreement with models about stellar wind interactions in the direct vicinity of Sgr~A*. For this, we assumed a stellar wind velocity of 750 km/s. We find several indications, that the observed Br$\gamma$ feature is influenced by wind and shock interactions with the ISM west of Sgr~A*. From the presented analysis, we suggest that the feature shows infrared line emission associated with a radio Bremsstrahlung filament. We can conclude, that we need more data that covers the area north and west of Sgr~A* to investigate the full size of the feature in the NIR or additional gas reservoirs. Since we also observe another prominent Br$\gamma$-bar L1 west of Sgr~A* (prominent in the MIR continuum emission), we are confident that observations with the JWST and future instruments like ERIS at the VLT will produce a more complete picture about the line emission features and filaments in the vicinity of Sgr~A*. It is possible that the gas reservoir associated with the extended feature can be connected to the dusty objects also present in this region. They could be indicators of ongoing star formation in the central region of the stellar cluster harboring Sgr~A*. Long term observations will complete this speculative connection.

\acknowledgments
We thank the referee for useful and constructive comments that helped to improve the manuscript.
This work was supported in part by the
Deutsche Forschungsgemeinschaft (DFG) via the Cologne
Bonn Graduate School (BCGS) and the Max Planck Society
through the International Max Planck Research School
(IMPRS) for Astronomy and Astrophysics. The Study of the conditions for star formation in nearby AGN
and QSOs is carried out within the Collaborative Research Centre 956,
sub-project [A02], funded by the Deutsche
Forschungsgemeinschaft (DFG) – project ID 184018867. Part of this
work was supported by fruitful discussions with members of
the European Union funded COST Action MP0905: Black
Holes in a Violent Universe and the Czech Science Foundation
-- DFG collaboration (No.\ 19-01137J). MZ acknowledges the financial support by the National Science Centre, Poland, grant No.~2017/26/A/ST9/00756 (Maestro 9). Harshitha Bhat received financial support for this research from the
International Max Planck Research School (IMPRS) for Astronomy and Astrophysics
at the Universities of Bonn and Cologne. We also would like to 
thank the members of the SINFONI/NACO/VISIR and ESO's Paranal/Chile team for their support and collaboration.


\appendix

\section{Proper motion measurements and individual blobs}

\label{sec:appendix}
As described in Sec. \ref{sec:propermotion}, we use different methods in order to derive a proper motion based on the prominent upper part.
\begin{figure}[ht!]
	\centering
	\includegraphics[width=1.\textwidth]{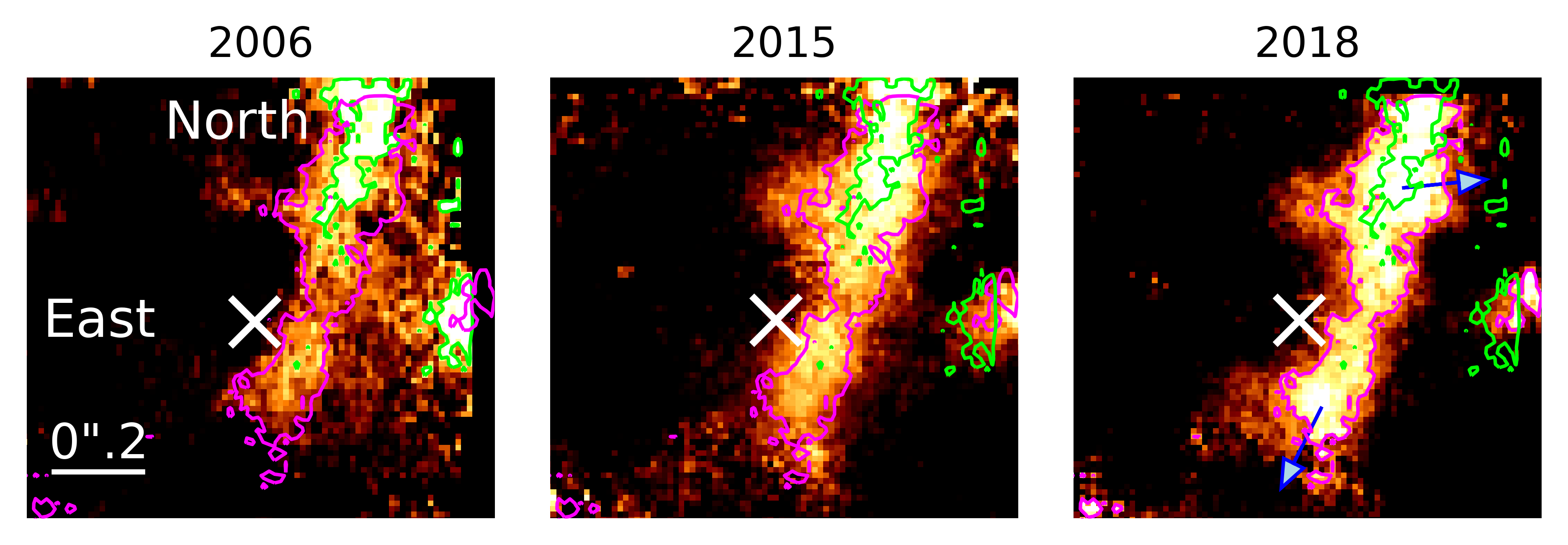}
	\caption{The central few arcseconds of the Galactic center in 2006, 2015, and 2018. The proper motion vector for the prominent upper part is pointing towards the west with a proper motion of around 320km/s. The southern part of the bar is moving towards the south-east with a proper motion velocity
of about 314km/s. The magenta contour lines corresponds to the 2018 line map, the green ones are based on the 2006 emission to demonstrate the evolution of the bar (45\% and 60\% of the related peak emission).}
\label{fig:pm}
\end{figure}
%The difference of the lower part between 2006 and 2015 is $20.47$mas, for the upper part $30.31$mas. With a distance of 8 kpc of the GC, the combined proper motion is around $106$km/s. This velocity is in line with other know feature properties. \cite{muzic2007} for example investigates thin dust filaments in the vicinity of Sgr~A*. The authors derive typical proper motions between $-100$km/s and $200$km/s. Figure \ref{fig:pm} show idications, that the B2V star S-star S2 blocks part of the Br$\gamma$-bar (see also Fig. \ref{fig:bar}). This can be noticed just above Sgr~A* in 2006. In 2018, when the star S2 is close to Sgr~A*, emission of the Br$\gamma$-bar in the blocked areas from 2006 can be observed. Next to the proper motion, this is another sign that the Br$\gamma$-bar is in the same plane as the S-cluster.
With these methods, we derive a proper motion of  around $320$km/s. This velocity is in line with other known features. \cite{muzic2007} for example investigate thin dust filaments in the vicinity of Sgr~A* and derive typical proper motions of several hundred km/s \citep[see also][]{Yusef-Zadeh1998, Zhao1998}. Figure \ref{fig:pm} shows indications that the B2V star S-star S2 blocks part of the Br$\gamma$-bar (see also Fig. \ref{fig:bar}). This can be noticed just above Sgr~A* in 2006. In 2018, when the star S2 is close to Sgr~A*, the emission of the Br$\gamma$-bar in the blocked areas from 2006 can be observed. %Next to the proper motion, this is another sign that the Br$\gamma$-bar is close to the S-cluster.

It should be noted that we see variable structures in the Br$\gamma$ line maps of the bar throughout the SINFONI data between 2006 and 2018 (Fig. \ref{fig:pm}). The less prominent lower part of the bar shows an increased intensity in 2018 compared to 2006. Considering the observation in 2015, the results indicate a stream of gas from north to south in contrast to the proper motion of the upper prominent part. This is supported by the lack of a bridge between the upper and lower part in 2006-2010 (see Fig. \ref{fig:pm}). At the position of the gap, we detect Br$\gamma$ emission in 2013-2018.

\section{Doppler-shifted line maps of the Br$\gamma$-bar L0}
Here, we show a line map overview. These channel maps are related to the emission lines discussed in Sec. \ref{sec:spectral_emission_lines}.
\begin{figure}[ht!]
	\centering
	\includegraphics[width=1.\textwidth]{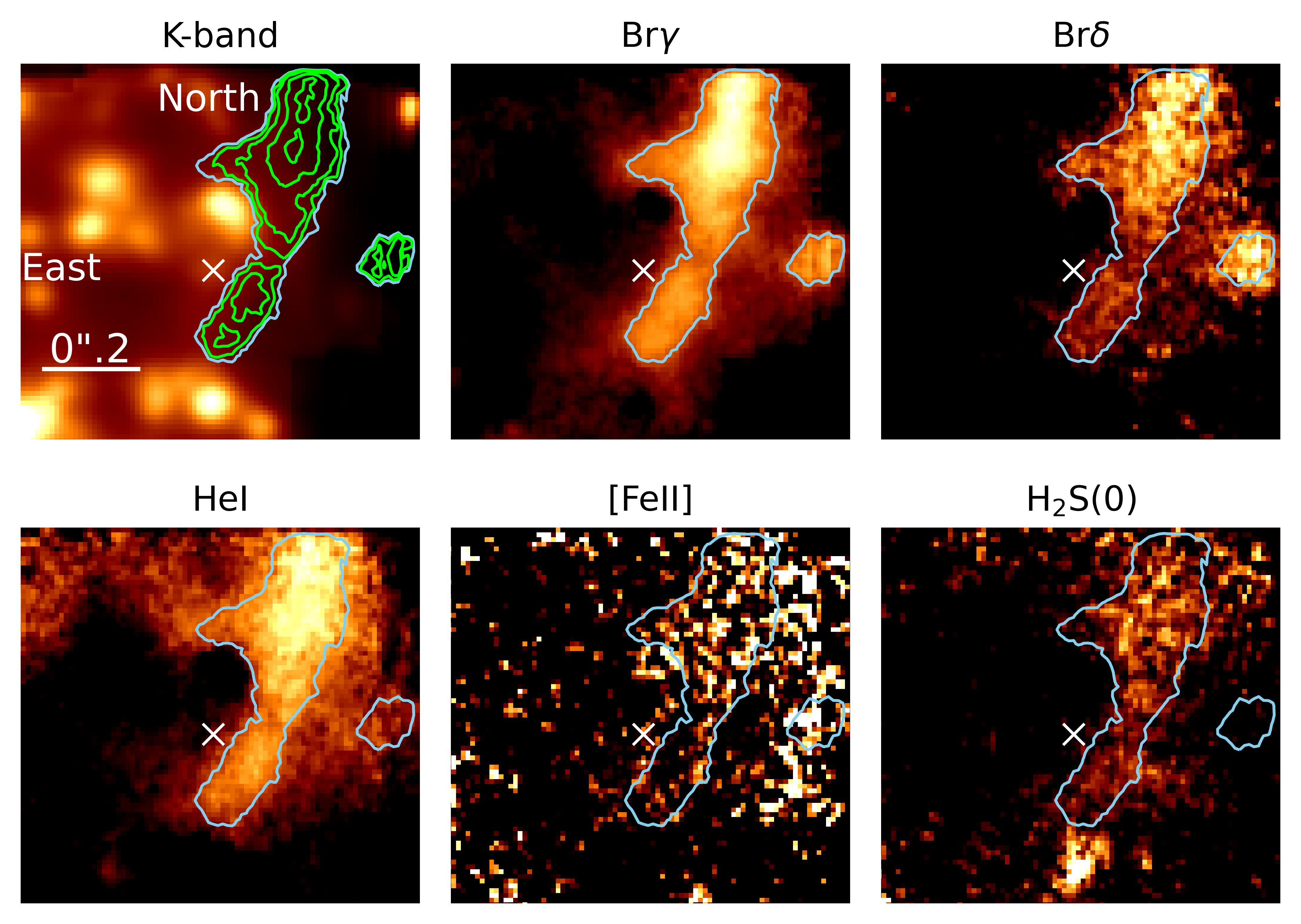}
	\caption{Line maps of the Br$\gamma$-bar L0. The upper left panel shows a K-band image of the GC based on the co-added data (see Fig. \ref{fig:bar}). For comparison, we add a 50\% contour line (sky-blue) to the channel maps.}
\label{fig:line_overview}
\end{figure}

\section{Data}
\label{sec:data_used}
In this section, we list the used SINFONI data between 2005 and 2018 (see Table \ref{tab:data_sinfo1}, \ref{tab:data_sinfo2}, \ref{tab:data_sinfo3}, \ref{tab:data_sinfo4}). We distinguish between medium and high quality that is measured by applying a Gaussian to the PSF of S2. Values for the FWHM in x and y direction that are higher than 7 px are considered medium, lower than 6.5 px are high. Because of a high airmass, bad seeing conditions or an insufficient AO correction, fitting a Gaussian to S2 is not always possible. We also exclude data, that is not following the object-sky-object pattern since the sky variability in the NIR shows an detectable impact on the data \citep{Davies2007}.
\begin{table*}[htbp!]
        \centering
        \begin{tabular}{cccccc}
        \hline\hline
        \\      Date & Observation ID  & \multicolumn{3}{c}{Amount of on source exposures} & Exp. Time \\  \cline{3-5} &  & Total & Medium & High &  \\
        (YYYY:MM:DD) &  &  &  &  & (s) \\ \hline\hline %table heading
        
        2005.06.16 & 075.B-0547(B) &  20  &  12   &  8  &    300  \\
        2005.06.18 & 075.B-0547(B) &  21  &  2   &  19  &    60  \\
        2006.03.17 & 076.B-0259(B) &  5  &  0   &  3  &    600  \\
        2006.03.20 & 076.B-0259(B) &   1 &   1  &   0 &    600  \\
        2006.03.21 & 076.B-0259(B) &  2  &   2  &  0  &    600  \\
        2006.04.22 & 077.B-0503(B) &  1  &   0  &  0  &    600  \\
        2006.08.17 & 077.B-0503(C) &  1  &   0  &  1  &    600  \\
        2006.08.18 & 077.B-0503(C) &  5  &   0  &  5  &    600  \\
        2006.09.15 & 077.B-0503(C) &  3  &   0  &  3  &    600  \\
        2007.03.26 & 078.B-0520(A) &  8  &   1  &  2  &    600  \\
        2007.04.22 & 179.B-0261(F) &  7  &   2  &  1  &    600 \\
        2007.04.23 & 179.B-0261(F) &  10 &   0  &  0  &    600 \\
        2007.07.22 & 179.B-0261(F) &  3  &   0  &  2  &    600  \\
        2007.07.24 & 179.B-0261(Z) &  7  &   0  &  7  &    600  \\
        2007.09.03 & 179.B-0261(K) & 11  &   1  &  5  &    600  \\
        2007.09.04 & 179.B-0261(K) &  9  &   0  &  0  &    600  \\
        2008.04.06 & 081.B-0568(A) &  16 &   0  &  15   &    600  \\
        2008.04.07 & 081.B-0568(A) &   4 &   0  &   4 &    600  \\
        2009.05.21 & 183.B-0100(B) &  7 &   0  &  7   &    600  \\
        2009.05.22 & 183.B-0100(B) &   4 &   0  &   4 &    400  \\
        2009.05.23 & 183.B-0100(B) &  2 &   0  &  2  &    400  \\
        2009.05.24 & 183.B-0100(B) &  3 &   0  &  3  &    600  \\
        
        \hline  \\
        \end{tabular}   
        \caption{SINFONI data of 2005, 2006, 2007, 2008, and 2009. The total amount of data is listed.}
        \label{tab:data_sinfo1}
        \end{table*}
\begin{table*}[htbp!]
        \centering
        \begin{tabular}{cccccc}
        \hline\hline
        \\      Date & Observation ID  & \multicolumn{3}{c}{Amount of on source exposures} & Exp. Time \\  \cline{3-5} &  & Total & Medium & High &  \\
        (YYYY:MM:DD) &  &  &  &  & (s) \\ \hline\hline %table heading
        
        2010.05.10 & 183.B-0100(O) & 3 &   0  &  3   &    600  \\
        2010.05.11 & 183.B-0100(O) &  5 &   0  &   5 &    600  \\
        2010.05.12 & 183.B-0100(O) & 13 &   0  &  13  &    600  \\
        2011.04.11 & 087.B-0117(I) &  3 &   0  &   3 &    600  \\
        2011.04.27 & 087.B-0117(I) & 10 &   1  &  9  &    600  \\
        2011.05.02 & 087.B-0117(I) & 6 &   0  &  6  &    600  \\
        2011.05.14 & 087.B-0117(I) & 2 &   0  &  2  &    600  \\
        2011.07.27 & 087.B-0117(J)/087.A-0081(B) & 2 &   1  &  1  &    600  \\     
        2012.03.18 & 288.B-5040(A) &  2 &   0  &   2 &    600  \\
        2012.05.05 & 087.B-0117(J) & 3 &   0  &  3  &    600  \\
        2012.05.20 & 087.B-0117(J) & 1 &   0  &  1  &    600  \\
        2012.06.30 & 288.B-5040(A) & 12 &   0  &  10  &    600  \\
        2012.07.01 & 288.B-5040(A) & 4 &   0  &  4  &    600  \\
        2012.07.08 & 288.B-5040(A)/089.B-0162(I)& 13 &   3  &  8  &    600  \\
        2012.09.08 & 087.B-0117(J)  & 2 &   1  &  1  &    600  \\
        2012.09.14 & 087.B-0117(J)  & 2 &   0  &  2  &    600  \\   
        2013.04.05 & 091.B-0088(A)  &  2 &   0  &   2 &    600  \\
        2013.04.06 & 091.B-0088(A)  & 8 &   0  &  8  &    600  \\
        2013.04.07 & 091.B-0088(A)  & 3 &   0  &  3  &    600  \\
        2013.04.08 & 091.B-0088(A)  & 9 &   0  &  6  &    600  \\
        2013.04.09 & 091.B-0088(A)  & 8 &   1  &  7  &    600  \\
        2013.04.10 & 091.B-0088(A)  & 3 &   0  &  3  &    600  \\
        2013.08.28 & 091.B-0088(B)  & 10 &  1  &  6  &    600 \\
        2013.08.29 & 091.B-0088(B)  & 7 &  2   &  4  &    600 \\
        2013.08.30 & 091.B-0088(B)  & 4 &  2   &  0  &    600 \\
        2013.08.31 & 091.B-0088(B)  & 6 &  0   &  4  &    600 \\
        2013.09.23 & 091.B-0086(A)  & 6 &   0  &  0  &    600  \\
        2013.09.25 & 091.B-0086(A)  & 2 &   1  &  0  &    600  \\
        2013.09.26 & 091.B-0086(A)  & 3 &   1  &  1  &    600  \\   
        
        \hline  \\
        \end{tabular}
        
        \caption{SINFONI data of 2010, 2011, 2012, and 2013.}
        \label{tab:data_sinfo2}
        \end{table*}

\begin{table*}[htbp!]
        \centering
        \begin{tabular}{cccccc}
        \hline\hline
        \\      Date & Observation ID  & \multicolumn{3}{c}{Amount of on source exposures} & Exp. Time \\  \cline{3-5} &  & Total & Medium & High &  \\
        (YYYY:MM:DD) &  &  &  &  & (s) \\ \hline\hline %table heading
        
        2014.02.27 & 092.B-0920(A) &  4 &   1  &  3  &    600   \\
        2014.02.28 & 091.B-0183(H) &  7 &   3  &  1  &    400   \\
        2014.03.01 & 091.B-0183(H) & 11 &   2  &  4  &    400  \\
        2014.03.02 & 091.B-0183(H) &  3 &   0  &  0  &    400   \\
        2014.03.11 & 092.B-0920(A) & 11 &   2  &  9  &    400   \\
        2014.03.12 & 092.B-0920(A) & 13 &   8  &  5  &    400   \\
        2014.03.26 & 092.B-0009(C) & 9  &   3  &  5  &    400   \\
        2014.03.27 & 092.B-0009(C) & 18 &   7  &  5  &    400   \\
        2014.04.02 & 093.B-0932(A) & 18 &   6  &  1  &    400    \\
        2014.04.03 & 093.B-0932(A) & 18 &   1  &  17 &    400    \\
        2014.04.04 & 093.B-0932(B) & 21 &   1  &  20 &    400       \\
        2014.04.06 & 093.B-0092(A) &  5 &   2  &  3  &    400      \\
        2014.04.08 & 093.B-0218(A) & 5  &   1  &  0  &    600   \\
        2014.04.09 & 093.B-0218(A) &  6 &   0  &  6  &    600   \\
        2014.04.10 & 093.B-0218(A) & 14 &   4  &  10  &    600   \\
        2014.05.08 & 093.B-0217(F) & 14 &   0  &  14  &    600   \\
        2014.05.09 & 093.B-0218(D) & 18 &   3  &  13  &    600   \\
        2014.06.09 & 093.B-0092(E) &  14 &   3  &  0  &    400   \\
        2014.06.10 & 092.B-0398(A)/093.B-0092(E) & 5 &   4  &  0   & 400/600 \\
        2014.07.08 & 092.B-0398(A)  & 6 &   1  &  3   &    600 \\
        2014.07.13 & 092.B-0398(A)  & 4 &   0  &  2   &    600 \\
        2014.07.18 & 092.B-0398(A)/093.B-0218(D)  & 1 &   0  &  0   &    600 \\
        2014.08.18 & 093.B-0218(D)  & 2 &   0  &  1   &    600 \\ 
        2014.08.26 & 093.B-0092(G)  &  4 &   3  &   0 &    400   \\
        2014.08.31 & 093.B-0218(B)  & 6 &   3   &   1 &    600 \\
        2014.09.07 & 093.B-0092(F)  & 2 &   0  &  0  &    400   \\
        2015.04.12 & 095.B-0036(A)  & 18 &  2 & 0 & 400 \\
        2015.04.13 & 095.B-0036(A)  & 13 &  7 & 0 & 400 \\
        2015.04.14 & 095.B-0036(A)  & 5  &  1 & 0 & 400 \\
        2015.04.15 & 095.B-0036(A)  & 23 &  13  & 10 & 400 \\
        2015.08.01 & 095.B-0036(C)  & 23 &   7  & 8  & 400 \\
        2015.09.05 & 095.B-0036(D)   17 &  11  & 4  & 400 \\

        \hline  \\
        \end{tabular}
        
        \caption{SINFONI data of 2014 and 2015.}
        \label{tab:data_sinfo3}
        \end{table*}
        
\begin{table*}[htbp!]
        \centering
        \begin{tabular}{cccccc}
        \hline\hline
        \\      Date & Observation ID  & \multicolumn{3}{c}{Amount of on source exposures} & Exp. Time \\  \cline{3-5} &  & Total & Medium & High &  \\
        (YYYY:MM:DD) &  &  &  &  & (s) \\ \hline\hline %table heading
        
        2018.02.13 & 299.B-5056(B) &  3 &   0  &  0  &    600   \\
        2018.02.14 & 299.B-5056(B) &  5 &   0  &  0  &    600   \\
        2018.02.15 & 299.B-5056(B) &  5 &   0  &  0  &    600   \\
        2018.02.16 & 299.B-5056(B) &  5 &   0  &  0  &    600   \\
        2018.03.23 & 598.B-0043(D) &  8 &   0  &  8  &    600   \\
        2018.03.24 & 598.B-0043(D) &  7 &   0  &  0  &    600   \\
        2018.03.25 & 598.B-0043(D) &  9 &   0  &  1  &    600   \\
        2018.03.26 & 598.B-0043(D) & 12 &   1  &  9  &    600  \\
        2018.04.09 & 0101.B-0195(B) &  8 &   0  &  4  &    600   \\
        2018.04.28 & 598.B-0043(E) & 10 &   1  &  1  &    600   \\
        2018.04.30 & 598.B-0043(E) & 11 &   1  &  4  &    600  \\
        2018.05.04 & 598.B-0043(E) & 17 &   0  &  17  &    600   \\
        2018.05.15 & 0101.B-0195(C) &  8 &   0  &  0  &    600   \\
        2018.05.17 & 0101.B-0195(C) &  8 &   0  &  4  &    600   \\
        2018.05.20 & 0101.B-0195(D) &  8 &   0  &  4  &    600   \\
        2018.05.28 & 0101.B-0195(E) &  8 &   3  &  1  &    600   \\
        2018.05.28 & 598.B-0043(F) &  4 &   0  &  4  &    600   \\
        2018.05.30 & 598.B-0043(F) &  8 &   5  &  3  &    600   \\
        2018.06.03 & 598.B-0043(F) &  8 &   0  &  8  &    600   \\
        2018.06.07 & 598.B-0043(F) & 14 &   1  &  7  &    600   \\
        2018.06.14 & 0101.B-0195(F) &  4 &   0  &  0  &    600   \\
        2018.06.23 & 0101.B-0195(F) &  8 &   1  &  1  &    600   \\
        2018.06.23 & 598.B-0043(G) &  7 &   2  &  1  &    600   \\
        2018.06.25 & 598.B-0043(G) & 22 &   5  &  7  &    600   \\
        2018.07.02 & 598.B-0043(G) &  3 &   0  &  0  &    600   \\
        2018.07.03 & 598.B-0043(G) & 22 &  12  & 10  &    600   \\
        2018.07.09 & 0101.B-0195(G) &  8 &   3  &  1  &    600   \\
        2018.07.24 & 598.B-0043(H) &  3 &   0  &  0  &    600   \\
        2018.07.28 & 598.B-0043(H) &  8 &   0  &  3  &    600   \\
        2018.08.03 & 598.B-0043(H) &  8 &   0  &  1  &    600   \\
        2018.08.06 & 598.B-0043(H) &  8 &   1  &  1  &    600   \\
        2018.08.19 & 598.B-0043(I) & 12 &   2  & 10  &    600   \\
        2018.08.20 & 598.B-0043(I) & 12 &   0  & 12  &    600   \\
        2018.09.03 & 598.B-0043(I) &  1 &   0  &  0  &    600   \\
        2018.09.27 & 598.B-0043(J) & 10 &   0  &  0  &    600   \\
        2018.09.28 & 598.B-0043(J) & 10 &   0  &  0  &    600   \\
        2018.09.29 & 598.B-0043(J) &  8 &   0  &  0  &    600   \\
        2018.10.16 & 2102.B-5003(A) &  3 &   0  &  0  &    600   \\

        \hline  \\
        \end{tabular}
        
        \caption{SINFONI data of 2018.}
        \label{tab:data_sinfo4}
        \end{table*}


\bibliography{bib}{}
\bibliographystyle{aasjournal}



\end{document}